# Finding Multiply Lensed and Binary Quasars in the DESI Legacy Imaging Surveys

C. Dawes[1], C. Storfer[2], X. Huang[3], G. Aldering[4], Aleksandar Cikota[5], Arjun Dey[6], and D. J. Schlegel[4]
[1] Department of Mathematics, Princeton University, Princeton, NJ 08544, USA; cdawes@princeton.edu
[2] Institute for Astronomy, University of Hawaii, Honolulu, HI 96822-1897, USA
[3] Department of Physics & Astronomy, University of San Francisco, San Francisco, CA 94117-1080, USA; xhuang22@usfca.edu
[4] Physics Division, Lawrence Berkeley National Laboratory, 1 Cyclotron Road, Berkeley, CA 94720, USA
[5] Gemini Observatory / NSF's NOIRLab, Casilla 603, La Serena, Chile
[6] NSF's National Optical-Infrared Astronomy Research Laboratory, 950 N. Cherry Avenue, Tucson, AZ 85719, USA

## Abstract

The time delay between multiple images of strongly lensed quasars is a powerful tool for measuring the Hubble constant ($H_0$). To achieve $H_0$ measurements with higher precision and accuracy using the time delay, it is crucial to expand the sample of lensed quasars. We conduct a search for strongly lensed quasars in the Dark Energy Spectroscopic Instrument (DESI) Legacy Imaging Surveys. The DESI Legacy Surveys comprise 19,000 deg$^2$ of the extragalactic sky observed in three optical bands ($g$, $r$, and $z$), making it well suited for the discovery of new strongly lensed quasars. We apply an autocorrelation algorithm to ~5 million objects classified as quasars in the DESI Quasar Sample. These systems are visually inspected and ranked. Here, we present 436 new multiply lensed and binary quasar candidates, 65 of which have redshifts from Sloan Digital Sky Survey Data Release 16. We provide redshifts for an additional 18 candidates from the SuperNova Integral Field Spectrograph.

## 1. Introduction

Multiply lensed quasars are powerful cosmological probes. In particular, quantifying the time delay between quasar images can deliver an independent measurement of the Hubble constant, $H_0$ (e.g., Refsdal 1964; Treu & Marshall 2016; Suyu et al. 2017). As tension persists between the value of $H_0$ inferred from the cosmic microwave background observations assuming the flat ΛCDM model (Planck Collaboration et al. 2020) and direct late-Universe measurements of $H_0$ (e.g., Riess et al. 2019, 2022), the development of a precise, accurate, and independent measurement of $H_0$ becomes ever more important. Careful analysis of small samples of lensed quasars have already yielded measurements of $H_0$ with competitive precision (e.g., Wong et al. 2020). For lensed quasars to make significant progress toward resolving the current tension, statistical uncertainties need to be reduced to below 1% (Treu & Marshall 2016), while effectively addressing systematic uncertainties (Birrer et al. 2020; Shajib et al. 2020). A larger and more diverse sample of lensed quasars can improve measurements of $H_0$ to subpercent precision (Rathna Kumar et al. 2015; Treu & Marshall 2016; Birrer & Treu 2021).

Beyond the study of cosmological parameters, lensed quasars offer opportunities for advancement in black hole physics, especially in understanding the co-evolution of supermassive black holes and their host galaxies. Local Universe ($z < 0.1$) observations have found that supermassive black hole mass ($M_{\rm BH}$) and various properties of the host galaxy (e.g., bulge luminosity, stellar velocity dispersion, and stellar mass) are tightly correlated, suggesting that central black holes and their host galaxies are physically coupled (Kormendy & Richstone 1995; Magorrian et al. 1998; Ferrarese & Merritt 2000; Gebhardt et al. 2000; Marconi & Hunt 2003; Häring & Rix 2004). To understand the coupling of central black holes and their host galaxies at earlier epochs, it is important to observe high-redshift galaxies beyond the local Universe. Strong gravitational lensing amplifies high-redshift quasars and magnifies their host galaxies. This, coupled with high-resolution imaging, allows for analysis of quasar and host galaxy properties through modeling and reconstruction (Ding et al. 2017, 2021). Despite this potential for lensed quasars in black hole physics, the relatively small number of known systems limits their impact, making the discovery of more lensed quasars important.

There are now ~200 confirmed lensed quasar systems (e.g., Weymann et al. 1979; Inada et al. 2012; More et al. 2016; Agnello et al. 2018; Lemon et al. 2018, 2019; Jaelani et al. 2021) and ~130 lensed quasar candidates (e.g., Sergeyev et al. 2016; Agnello et al. 2018; Chan et al. 2020), a development that can be mostly attributed to large-scale surveys such as the Sloan Digital Sky Survey (SDSS; e.g., Inada et al. 2003), the Kilo-Degree Survey (e.g., Spiniello et al. 2018), the Dark Energy Survey (DES; e.g., Anguita et al. 2018), PanSTARRS (e.g., Rusu et al. 2019), and Gaia (e.g., Lemon et al. 2018). As the search for lensed quasars pivots toward large-scale surveys, recent papers tend to include many candidates alongside new confirmed systems (e.g., Chan et al. 2020; Lemon et al. 2020). This trend can be attributed to the much increased size of the new large-scale surveys and high volume of new discoveries. Upon further examination of these candidates, many quasar pairs, in particular, turn out to be physical quasar–quasar binaries rather than the result of lensing (Kochanek et al. 1999;

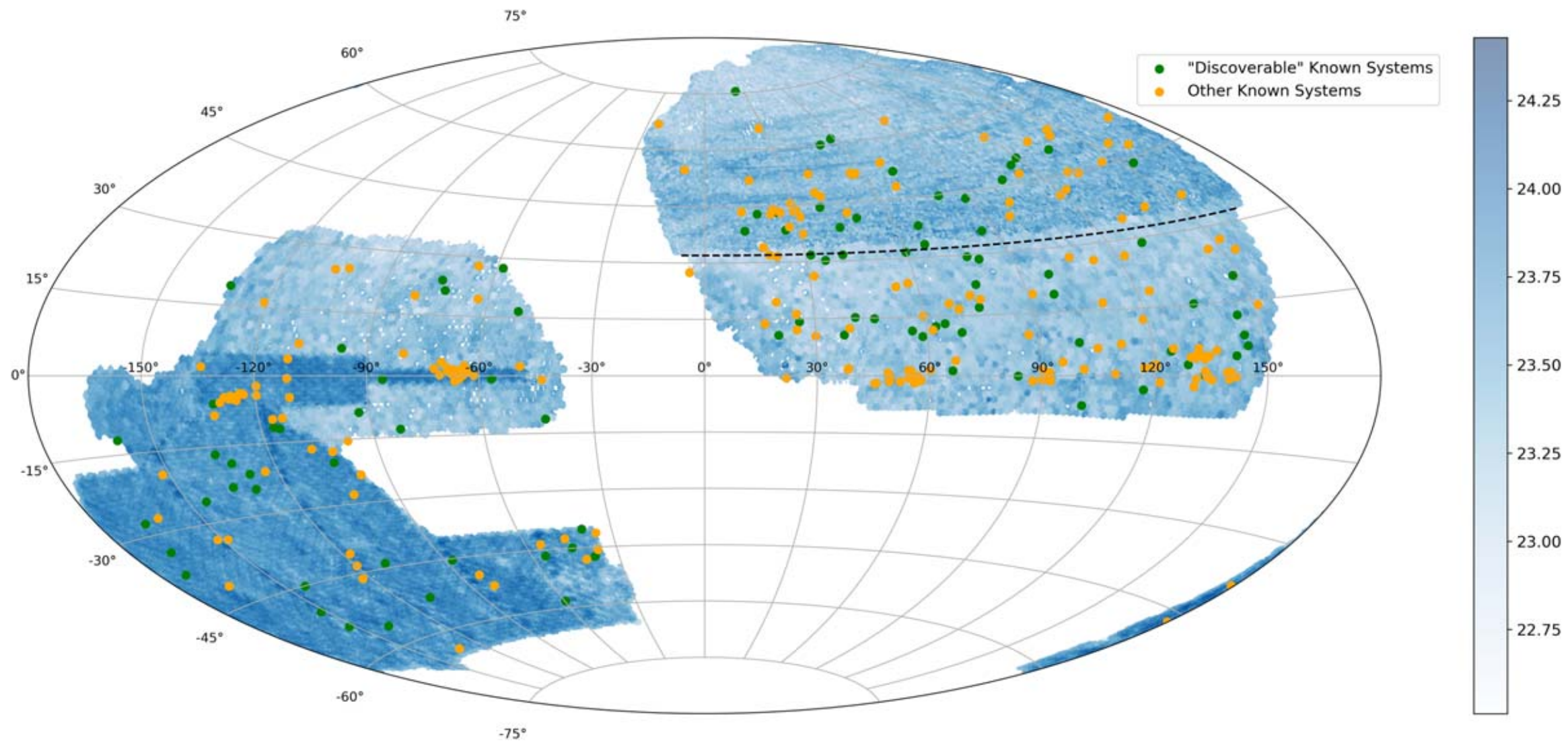

**Figure 1.** The DESI Legacy Imaging Surveys footprint in an equal-area Aitoff projection in equatorial coordinates. The dotted line separates the north (MzLS/BASS) and south (DECaLS) subregions. Slightly above $\delta = 32°$, there is a small amount of overlap between MzLS/BASS and DECaLS. Patches with different shades of blue indicate the depth in the $z$ band ($5\sigma$ detection significance). "Discoverable" lensed quasars and other known lensed quasars (see Section 3, Table 1) within the Legacy Surveys footprint are overlaid.

Mortlock et al. 1999). Quasar pairs that turn out to be physical binaries are also valuable, however, as physical binaries lend insight to how galaxy–galaxy interactions and mergers can enhance or even trigger quasar activity (Begelman et al. 1980; Djorgovski 1991; Di Matteo et al. 2005; Hopkins et al. 2008; Ellison et al. 2011; Liu et al. 2012; Bogdanović et al. 2022), or as the progenitor of binary black hole systems (e.g., Boroson & Lauer 2009).

We have searched the Dark Energy Spectroscopic Instrument (DESI) Legacy Imaging Surveys for multiply lensed quasars. Previous papers have searched the DESI Legacy Imaging Surveys for strongly lensed galaxies (Huang et al. 2020, 2021; Stein et al. 2022; Storfer et al. 2022), but this is the first search of the survey for lensed quasars in particular. We note that in this paper, "strongly lensed" and "multiply lensed" are used interchangeably. In Section 2, we give an overview of the DESI Legacy Surveys. We describe our methodology in Section 3. In Section 4, we present the results of our search, as well as the spectroscopic observations of a subset of our candidates. We discuss our results in Section 5 and conclude in Section 6.

## 2. Observations

The DESI Legacy Imaging Surveys (Dey et al. 2019) cover $\sim$19,000 deg$^2$ of the extragalactic sky visible from the northern hemisphere. The footprint is split into two contiguous parts by the Galactic plane, and each was observed with at least three passes in the $grz$ bands (Figure 1). The Legacy Surveys are composed of a northern region and a southern region (see Figure 1). The 14,000 deg$^2$ southern region, with $\delta \lesssim +32°$, comprises the Dark Energy Camera Legacy Survey (DECaLS). For the 5000 deg$^2$ northern region, the Beijing–Arizona Sky Survey (BASS; Zou et al. 2017) carried out the $gr$-band observations and the Mayall $z$-band Legacy Survey (MzLS), the $z$-band observations. For DECaLS, the $g$, $r$, and $z$ bands deliver an image quality with FWHM of approximately 1$''$29, 1$''$18, and 1$''$11, respectively, compared with 1$''$61, 1$''$47, and 1$''$01 in the MzLS/BASS subregion. Figure 1 shows the subregions of the Legacy Survey footprint and their depths.

The Legacy Surveys serve as the precursor to the DESI spectroscopic survey with the purpose of identifying targets for spectroscopic observations. The Legacy Surveys source catalogs are constructed using *The Tractor* (Lang et al. 2016), a forward-modeling algorithm that performs source extraction on pixel-level data. A small set of light profiles (point source, point-spread function, or PSF; round exponential galaxy with a variable radius; de Vaucouleurs profile, for elliptical galaxies; exponential profile, for spiral galaxies; and Sérsic profile) are fit to each source to determine the best-fit model.[7] Important to our search are sources modeled as PSFs, which are likely quasars or stars.

The DESI Quasar Sample (Yèche et al. 2020) identifies the potential quasars among DESI targets. DESI quasar selection utilized the three optical bands ($grz$) of the Legacy Surveys as well as the W1 and W2 bands of the Wide-field Infrared Survey Explorer (WISE; Wright et al. 2010). The criteria for selection are restricted to objects with stellar morphology (PSFs), $r$-band magnitude <22.7 AB mag, and to targets not in areas with corrupted or masked pixels (e.g., targets in the vicinity of bright stars, globular clusters, or bright galaxies). Yèche et al. (2020) used two distinct methods for quasar target selection: color cuts and a machine-learning algorithm. The color cuts consist of W1 − W2 and $r$ − W versus $g - z$, where W is the magnitude computed from the weighted average of W1 and W2 fluxes. The W1 and W2 bands are crucially important in distinguishing quasars from stars due to the infrared excess observed in quasars at all redshifts. In addition, a random forests (RF)–based algorithm has been trained on 230,000 known quasars within the Legacy Surveys footprint and 210,000 unclassified objects presumed to be stars. 98% of quasar targets selected via the color-cut method are contained within the RF selection, which is more complete than the color-cut method at low and high redshifts.

[7] For more details, see https://www.legacysurvey.org/dr9/description/.

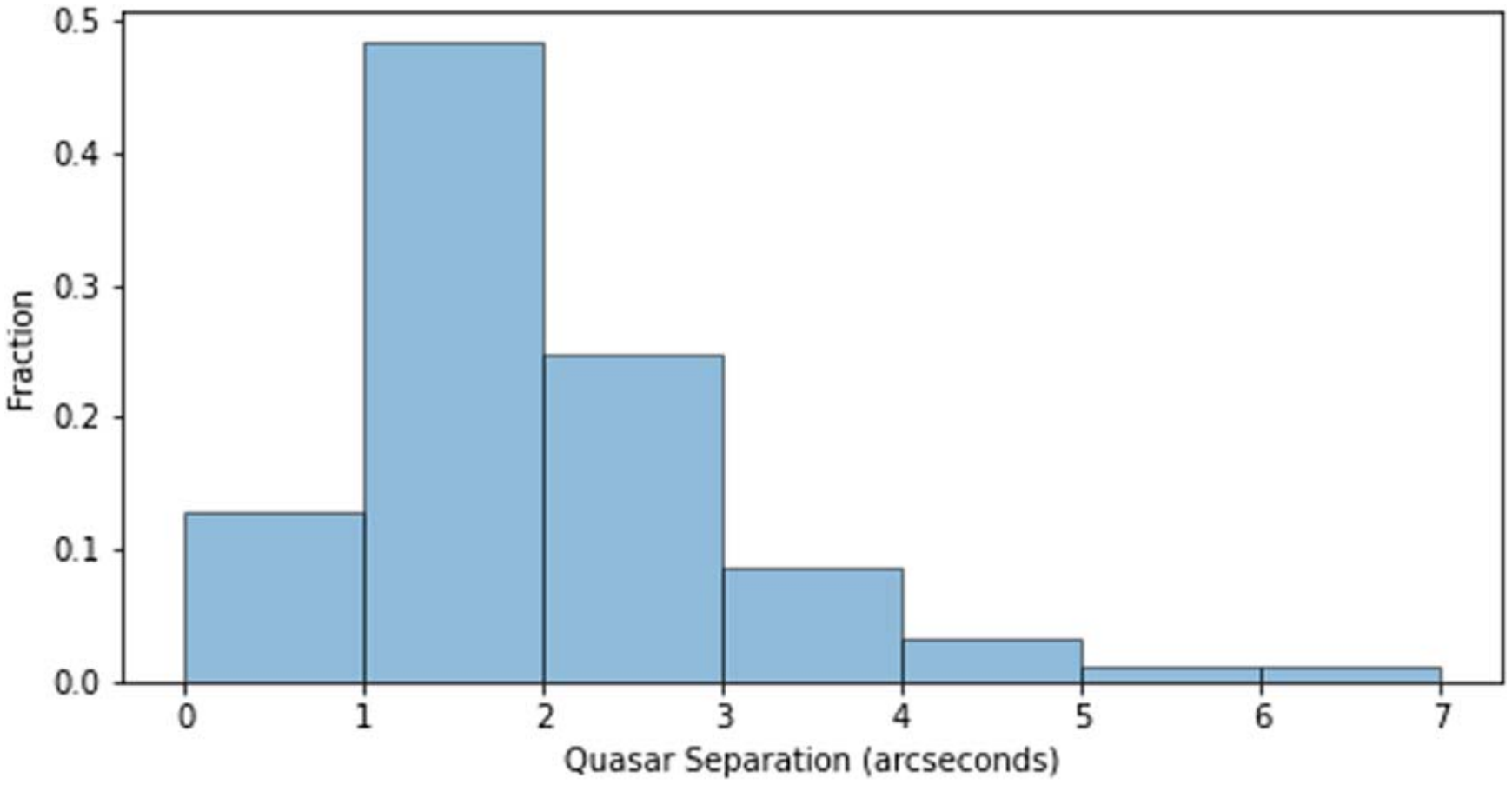

**Figure 2.** Histogram of separations between lensed quasar images for the 94 "discoverable known" systems (see the text).

**Table 1**
Lensed Quasar System Types

| Name | Description | Number of Systems |
|---|---|---|
| Known | Known lensed quasar systems and candidates within the DESI Legacy Surveys footprint | ∼300 |
| "Discoverable Known" | Among the ∼300 known systems, those that are discoverable by our algorithm (see Section 3) | 94 |
| "Discoverable Confirmed" | Those "discoverable known" systems that are confirmed | 72 |
| New Candidates | New candidate multiply lensed and binary quasar systems | 436 |

While both methods have been employed in selecting targets for the DESI main survey, the current DESI Quasar Sample corresponds to the RF selection.

## 3. Methodology

We assemble a catalog of known lensed quasars within the Legacy Surveys footprint, composed primarily of three separate, earlier catalogs: the Master Lens Database (Moustakas et al. 2012), the Gravitationally Lensed Quasar Database,[8] and the VizieR Online Data Catalog: Gaia GraL. II. Known Multiply Imaged Quasars (Delchambre et al. 2019). We have since added dozens of confirmed lenses and candidates from more recent publications (Sonnenfeld et al. 2018, 2020; Chan et al. 2020; Jaelani et al. 2021; Stern et al. 2021). Several of these papers present both candidates and confirmed systems (e.g., Agnello et al. 2018). In total, the catalog consists of ∼300 known lensed quasars and candidates (see Table 1). The compilation of this known lens catalog is used to inform decisions regarding our autocorrelation algorithm and subsequent human inspection.

To search for lensed quasars in the DESI Legacy Surveys, we perform autocorrelation on the aforementioned DESI Quasar Sample, which contains over 5 million targets. About 30% (94) of the ∼300 known lensed quasars within the Legacy Surveys footprint have two or more objects in the DESI Quasar Sample and therefore are "discoverable" by our algorithm (henceforth, "discoverable known" systems; see Figure 1, Table 1). Figure 2 shows the distribution of image separations among "discoverable known" systems. For the remaining known systems (∼70%), a subset are the candidates from the SuGOHI project (Sonnenfeld et al. 2018, 2020; Chan et al. 2020; Jaelani et al. 2021), which, though promising lensed candidates, appear qualitatively different from the other known systems in the Legacy Surveys.[9] Excluding the SuGOHI candidates, the distribution of image separations for the rest[10] is similar to the 94 "discoverable known."

To err on the side of completeness, we identify and group multiple quasar targets within a 10″ radius of each other. We later reduce our separation cut by half, because more than 95% of the "discoverable known" systems have quasar targets separated by less than 5″. We perform autocorrelation within each of the Legacy Survey "bricks." "Bricks" are approximately $0.26 \times 0.26\,\mathrm{deg}^2$ (3600 pixels × 3600 pixels) in size and typically contain 15–20 quasar targets each. Performing autocorrelation within each brick risks missing groupings of quasars that straddle the boundaries of two or more bricks. With a 5″ image separation cut, we would miss some systems whose center happens to lie within 2.″5 of a brick boundary. However, only some such systems are lost, depending on their position angle. Accounting for position angle, 0.3% is a generous upper limit to the fraction of lensed quasar candidates we would have missed due to not searching for a nearby target in a neighboring brick.

We developed our own algorithm in Python. The algorithm proceeds brick by brick (we made use of the mpi4py package to run in parallel across bricks). In each brick, the algorithm first groups pairs of quasars within 10″ of each other and then recursively connects multiple pairs of quasar targets within the same candidate system (such as would occur in "triples" and "quads").[11] We test our algorithm on the 94 "discoverable known" systems and simulations, achieving a 100% success rate. We then deploy our autocorrelation algorithm on the ~5 million quasar targets across the Legacy Surveys footprint.

[8] https://research.ast.cam.ac.uk/lensedquasars/

[9] For example, for Chan et al. (2020), even though the title mentions the search for quasars, their Figure 1 shows that many of their candidates are probably galaxy–galaxy lensing systems.

[10] Note that even for these, multiple quasars were not identified in the DESI Quasar Sample (hence these are among the non-"discoverable"). However, this is not indicative of the overall completeness of the DESI Quasar Sample, which relies on WISE data with a resolution of 6″, whereas we are looking for quasars <5″ apart.

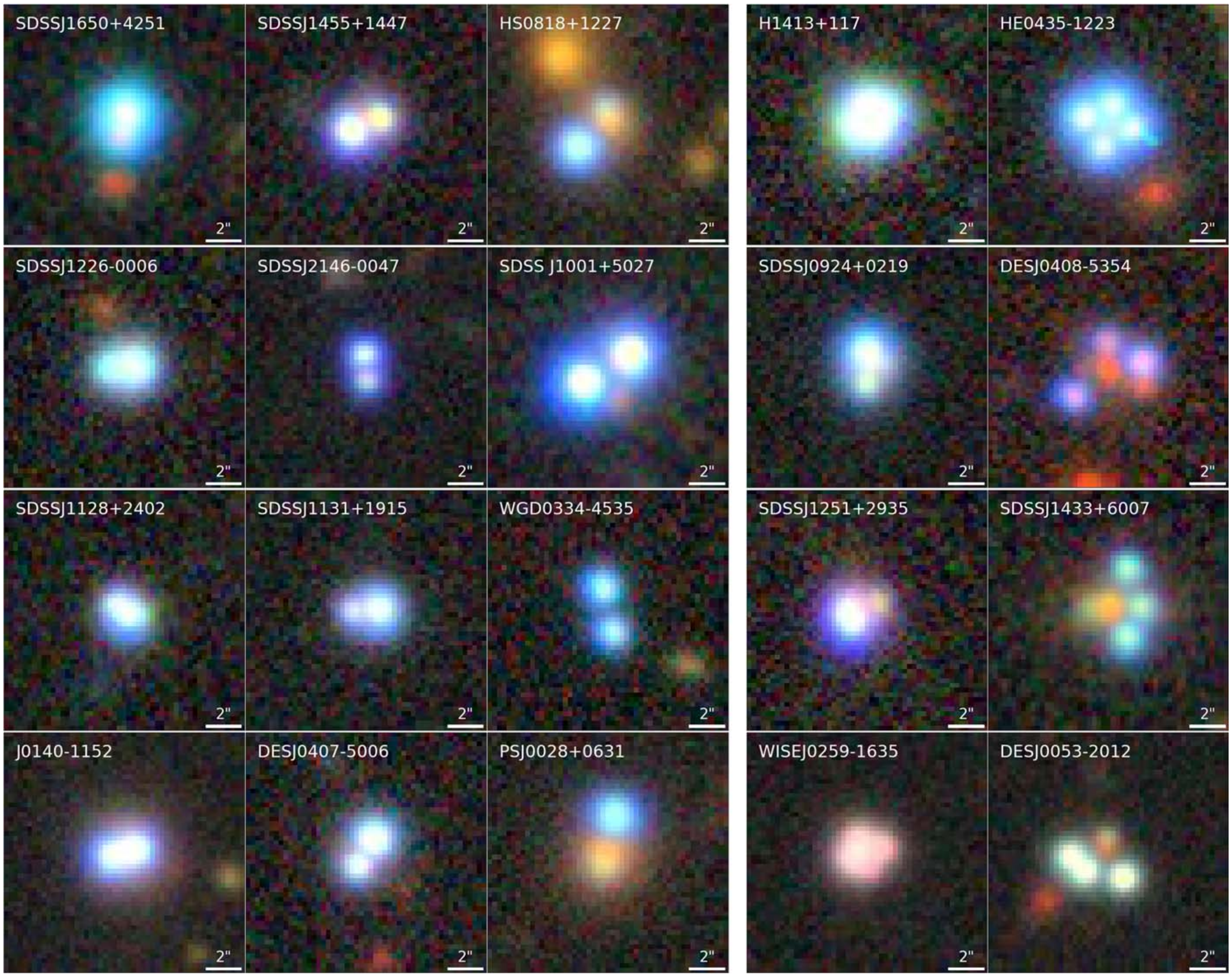

**Figure 3.** In this figure, we show representative configurations of the "discoverable confirmed" lensed quasar systems. The left panel consists of doubles: the first column shows systems that require deblending; the second column shows small-image-separation systems ($<2''.5$); and the third column shows large-image-separation systems ($\gtrsim 2''.5$). The right panel consists of quads: the first column shows systems that require deblending and the second column shows systems that do not.

## 4. Results

In this section, we present the results from the autocorrelation and subsequent human inspection. The $10''$ radius search found over 27,000 recommendations (quasar targets, each of which is within $10''$ of at least one other). Due to the abundance of recommendations and because more than 95% of the aforementioned "discoverable known" lenses (see Table 1) have quasar targets that are separated by less than $5''$, we choose to impose a cut on the recommendations to those with separation $<5''$. This results in almost 6000 recommendations. Based on this and the DESI Quasar Sample surface density, we estimate that 2.4 recommendations will be chance superpositions. These will likely make it into our sample.

These recommendations are subjected to human inspection and graded as A, B, C, or nonlens. Of the 94 "discoverable known" systems, 72 are confirmed (henceforth, "discoverable confirmed" systems; see Table 1). The grading criteria are informed by the appearance of these "discoverable confirmed" systems in Legacy Surveys images. Figure 3 shows an assortment of these confirmed lenses. The following are common features of known lensed quasars that guided the visual inspection and grading:

1. quasar targets with similar color that are $<2''.5$ apart, but sometimes slightly farther;
2. the presence or hint of a redder (putative lensing) galaxy between the quasar images;
3. in the case of unclear lensing galaxy light, one of the quasar images appearing dimmer and/or redder, which may be due to its closer proximity to the lensing galaxy;

[11] Here by "triple" or "quad," we only mean that three or four close-by images are identified in the DESI Quasar Sample, while at this point, we remain agnostic about the lensing configuration.

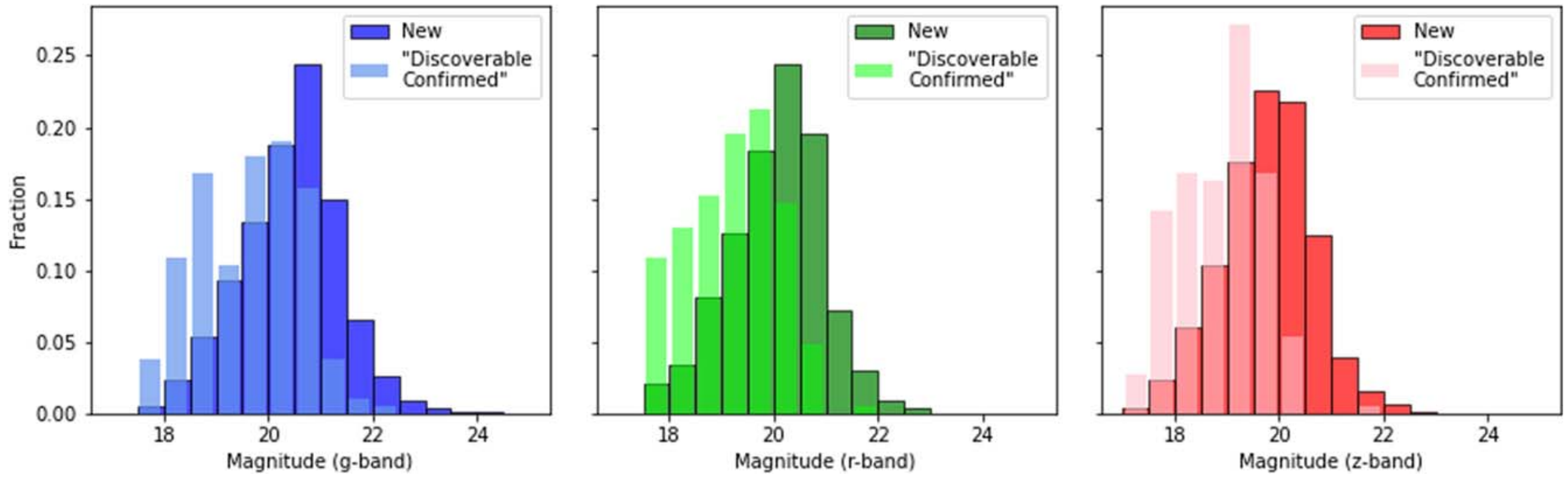

**Figure 4.** Distributions of *g*-, *r*-, and *z*-band magnitudes of the images of new candidates and "discoverable confirmed" systems. Note that the distributions of our candidates are fainter than the "discoverable confirmed" systems.

4. in the case of close pairs/quads (image separation comparable to or only slightly less than the seeing), two or more quasar targets deblended by *The Tractor* appearing as an elongated bright object; and
5. four quasar images arranged in an Einstein Cross–like configuration around a lensing galaxy.

Because lensed quasars typically display a combination of these features, there is great diversity in their appearance. Furthermore, for some of the candidate systems, *The Tractor* has identified additional point sources, providing evidence that some systems we initially thought were doubles are actually quads (for 17 of the 24 quads identified by our algorithm, they each have only two objects from the DESI Quasar Sample, but three or four point sources from *The Tractor*).

To begin, C.D. and C.S. made "first pass" selections according to the above criteria. As a "second pass," C.D., C. S., and X.H. together assigned an integer score between 1 and 4 to the "first pass" selections, while coauthor A.D. did the same independently. These two scores were then averaged. This grading scheme is similar to that of Huang et al. (2021). The final grades for the candidates can be broken down as follows:

1. $\geqslant 3.5$: Grade A. We are highly confident that these candidates are lensed quasars. Many of them have angular separations larger than the seeing and the putative lens is visible (or there is a hint of lens light) and the quasar images have similar colors. Others are close pairs and quads.
2. $= 3.0$: Grade B. These have characteristics that are similar to those of the Grade As, but weaker. The lens light is often not obvious in close pairs. For some systems, the putative counterimages are somewhat redder, possibly due to contamination from the lens light, or just fainter.
3. $= 2.5$ or 2.0: Grade C. These systems generally have large image separations and are fainter than the As and Bs. Many of them do not have a discernible lensing galaxy. For the few cases with a possible lens or lens light, they have an atypical lensing configuration.

In total, we have identified 530 candidate lensed quasar systems. As stated before, all 94 "discoverable known" systems have been rediscovered by our algorithm as recommendations. We therefore have found 436 new candidate systems. Table 1 gives a breakdown of the relevant types of systems discussed in this paper.

Figure 4 shows the magnitude distribution of our newly discovered candidates compared to that of "discoverable confirmed" systems. Our candidates include many more faint objects than the confirmed lensed quasars found in previous surveys. Figure 5 shows our new candidates plotted over a depth map of the Legacy Surveys. Among the new candidates, there are 102 As, 118 Bs, and 216 Cs. Figure 6 shows 10 notable systems, with nine doubles and one potential quad.

Figure 7 and Table 2 show the first 80 of our new candidates by ascending R.A., grouped by grade. All 436 new candidates can be found on our project website: https://sites.google.com/usfca.edu/neuralens.

### 4.1. Gaia Proper Motions and Parallax

Gaia Early Data Release 3 is used to provide further checks on the quality of our candidates; specifically, the low significance of both proper motion (PM) and parallax is an important indicator that a candidate system is composed of quasars and not Milky Way stars. 380 ($\sim$87%) of our candidates have Gaia PM and parallax information for at least one quasar image, which we obtained from the Gaia Archive.[12] We follow a different definition of PM significance (PMSIG) compared to Lemon et al. (2019). By our definition,

$$\mathrm{PMSIG} = \sqrt{\frac{\mathrm{PM_RA}^2 + \mathrm{PM_DEC}^2}{\sigma^2_{\mathrm{PM_RA}} + \sigma^2_{\mathrm{PM_DEC}}}}.$$

Additionally, we take into consideration parallax significance (PXSIG), where

$$\mathrm{PXSIG} = \frac{\mathrm{PX}}{\sigma_{\mathrm{PX}}}.$$

Figure 8 shows the distributions of PMSIG and PXSIG for the 380 new candidates and 72 "discoverable confirmed" systems with Gaia information. Informed by the PMSIG and PXSIG of the "discoverable confirmed" systems, we decide that PMSIG $< 8$ and PXSIG $< 3.5$ are acceptable for our candidates.

### 4.2. Redshifts for Our Candidate Systems

For the vast majority of our systems, the putative lensing galaxy is much fainter than the lensed images. As a result, we are only able to find photometric redshifts for 27 lenses (Section 4.2.1). For the lensed sources (quasars), we identified 65 systems with spectroscopic redshifts from SDSS and we have additionally obtained spectroscopy using the SuperNova Integral Field Spectrograph (SNIFS; Aldering et al. 2002) for a subset of our systems (Section 4.2.2). Six of our candidates have photometric redshifts for the foreground galaxy and spectroscopic redshifts for the background quasar. These are presented in Section 4.2.3.

[12] https://gea.esac.esa.int/archive/

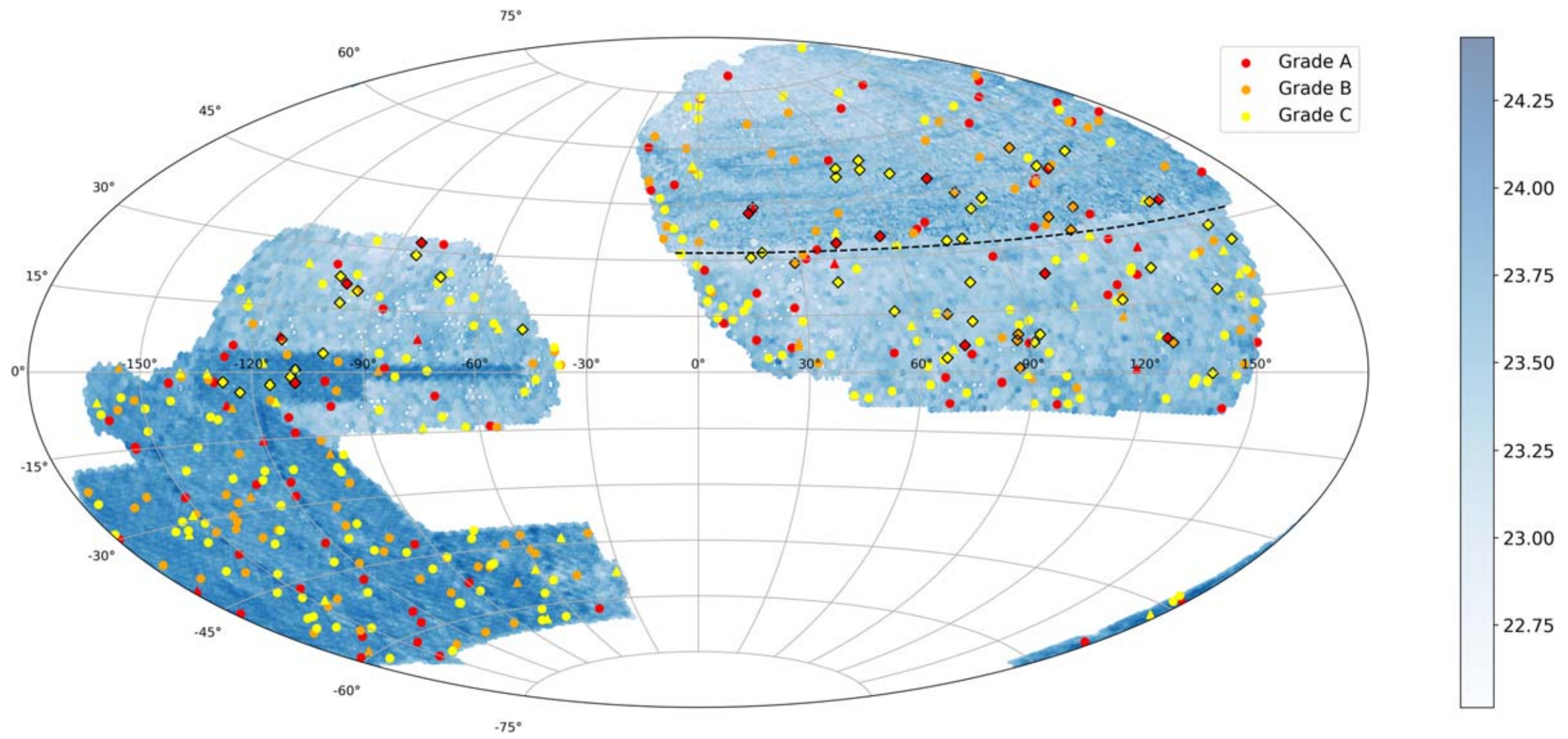

**Figure 5.** Grade A, B, and C candidates overlaid on the DESI Legacy Surveys (see Figure 1) are shown as red, orange, and yellow dots, respectively (those without Gaia data as triangles; see Section 4.1). Note that the 94 "discoverable known" systems are not included. Candidates with spectroscopic redshifts from SDSS DR16 (see Section 4.2.2) are outlined by a black diamond.

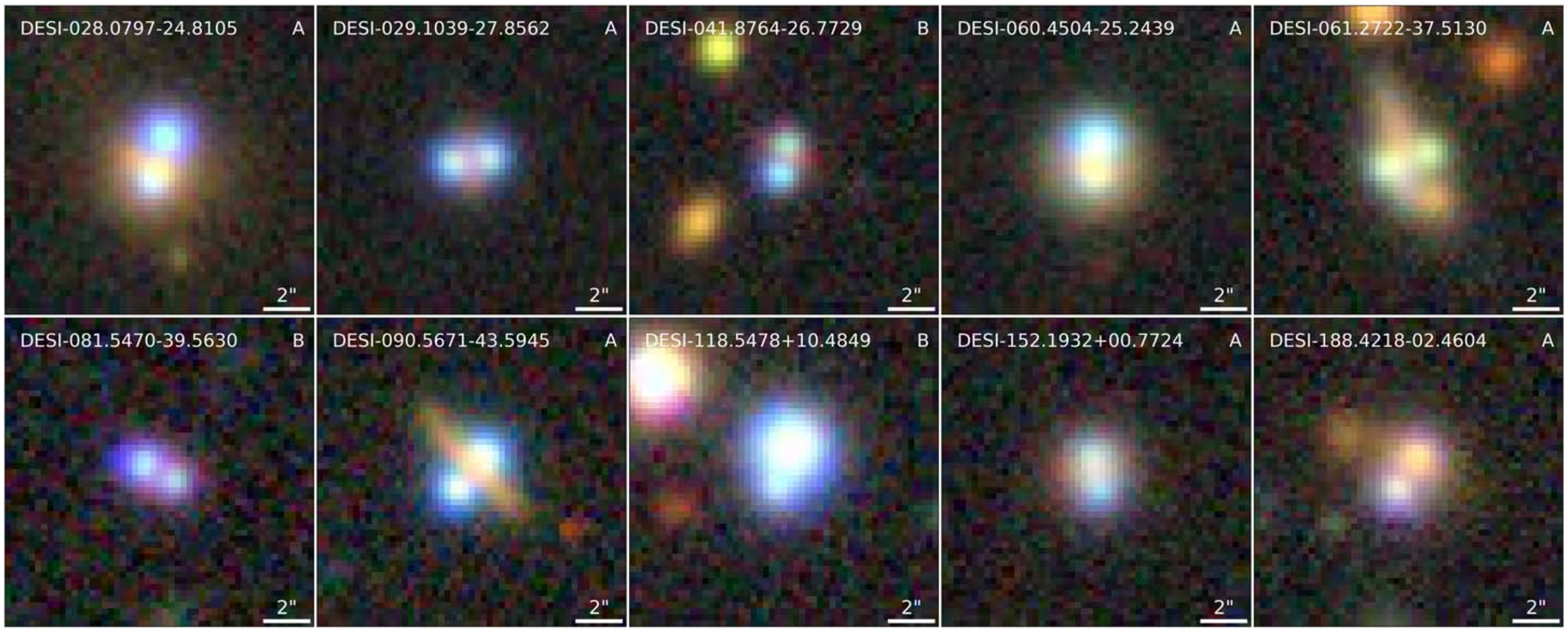

**Figure 6.** Ten visually impressive candidates. The naming convention is R.A. and decl. in decimal format. The top right corner of each image indicates its grade. North is up, and east is to the left. DESI-118.5478 + 10.4849 is a potential quad and the other nine are doubles. Note that observation by the blue channel of SNIFS did not detect quasar features in DESI-090.5672-43.5945, possibly due to the low S/N of the spectrum.

#### 4.2.1. Photometric Redshifts of Lensing Galaxies

For redshift information on our lensing galaxies, we include photometric redshifts for 27 of our candidate systems from Zhou et al. (2021). In cases where the centroid of *The Tractor*–extracted source is ambiguous (i.e., straddling the putative lens and a lensed image), we did not include the photometric redshift. In cases where the putative lens and at least one lensed image are very close, the photometric redshift may be less reliable.

#### 4.2.2. Spectroscopic Redshifts from SNIFS and SDSS

We have observed a subset of our high-quality candidates with $r$-band mag $\lesssim$20.0 on the SNIFS instrument on the University of Hawaii 88 inch (2.2 m) telescope (UH88), located on Maunakea. The spectrograph has two channels, blue (320 nm to 560 nm) and red (520 nm to 1$\mu$ m), with resolutions of $R$ ~1000 and $R$ ~1300, respectively. Our candidates were observed on the nights of 2020 September 21, October 12, October 19, and November 9 as well as 2021 June 11, June 14, July 12, and November 8 UTC. SNIFS has a $6.4 \times 6''.4$ field of view. Given this field of view, we do not need to impose an image separation cut for our targets. SNIFS splits its field of view via a $15 \times 15$ microlens array into 225 samples, each $0.43 \times 0\farcs43$ (Lantz et al. 2004). The spectra were extracted from each spatiospectral data cube using a circular aperture with a radius of $2\sigma$ and a surrounding annual sky region spanning 5–7$\sigma$, where $\sigma$ is the second moment of the PSF of the lensed quasar candidate system. For this pilot project, we did not attempt to disentangle the light from the individual sources.

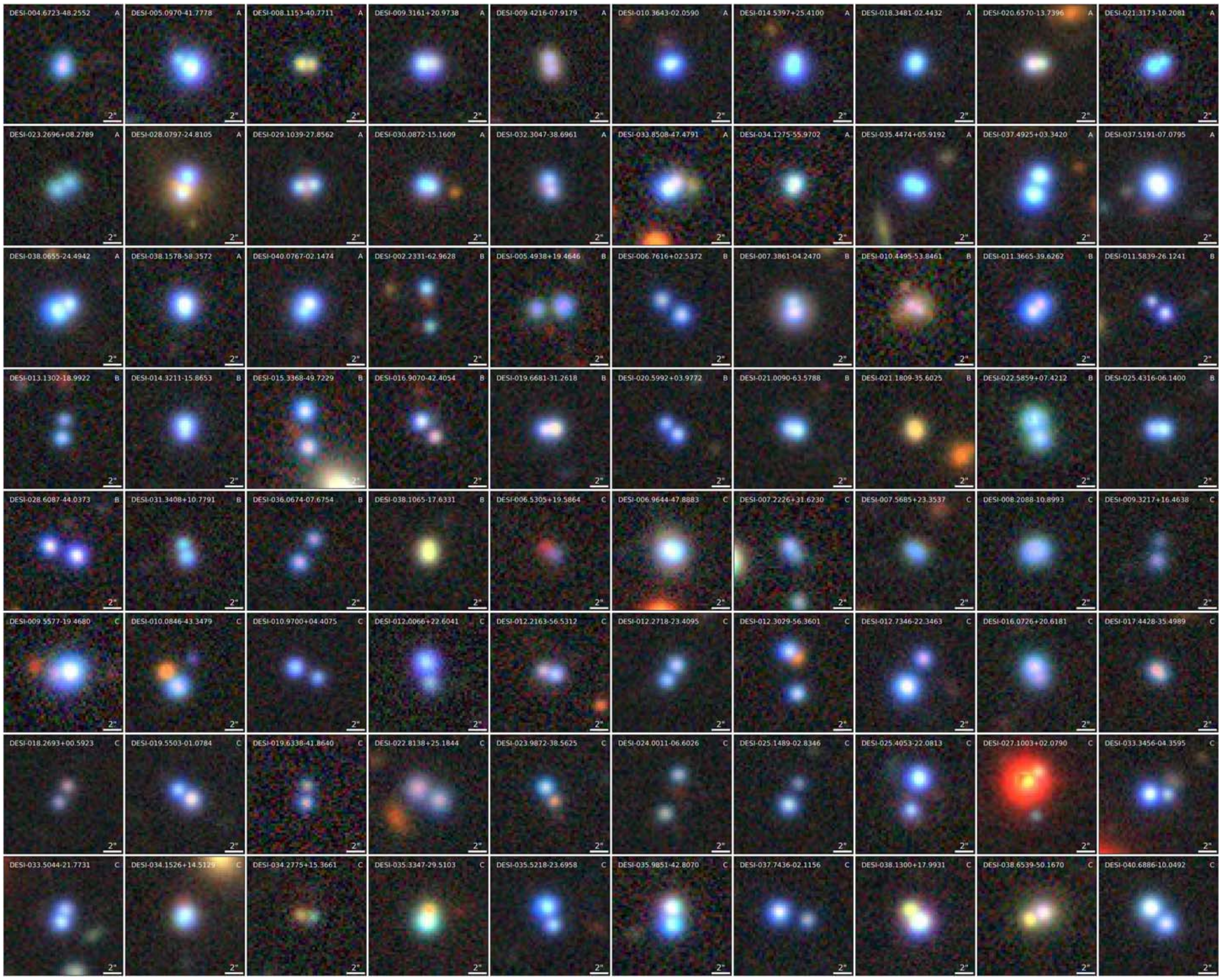

**Figure 7.** The first 80 of our newly discovered candidates by ascending R.A., grouped by grade. For the naming convention, see the Figure 6 caption. The top right corner of each image indicates its grade. North is up, and east to the left. Additional information about each system can be found in Table 2.

We fit for the SNIFS spectra using the publicly available SDSS Data Release (DR) 5 spectral template for quasars.[13] Of the 29 candidates observed, 10 had spectra with an insufficient signal-to-noise ratio (S/N; due to poor observing conditions), one was a Milky Way star,[14] and 18 were confirmed as quasars. For these 18, the best-fit redshifts range from $z = 0.75$ to 2.60 (Figure 9). We later found that DESI-237.7387 + 02.3629 was observed by SDSS with a redshift consistent with the value from SNIFS (both are included in the online table on our project website).

For additional redshift information, we also include spectroscopic redshifts from SDSS DR16 for 65 quasars in our candidate systems.[15] The redshifts of these quasars range from 0.36 to 3.55, and 61 of them have $z > 0.5$. Figure 11 shows redshift distributions for our candidates. The high-redshift distribution of the quasars is consistent with them being lensed.

We note that DESI-037.5191-07.0795 appears to be a quasar with broad absorption-line (BAL) activity.[16] We identify three potential rare quasar–quasar lensing systems: (1) DESI-033.3456-04.3595, with foreground and background redshifts of $(z_{fg},\ z_{bg}) = (1.9052,\ 3.5458)$ and an image separation of $1\farcs91$; (2) DESI-207.6277 + 52.4015, with $(z_{fg},\ z_{bg}) = (0.9747,$ $3.2027)$ and image separation $3\farcs37$; and (3) DESI-343.7205 + 23.8984, with $(z_{fg},\ z_{bg}) = (0.4866,\ 2.0103)$ and

[13] http://classic.sdss.org/dr5/algorithms/spectemplates/, Template 30.

[14] This was subsequently removed from our candidate list and is therefore not among the 436 candidates reported in this paper.

[15] For two systems with two images each observed by SDSS, the redshift difference for the two images in each system is inconsistent with lensing. These two were removed from our candidates.

[16] See http://classic.sdss.org/dr5/algorithms/spectemplates/, Template 32.

**Table 2**
Eighty of the New Candidates Arranged in Ascending R.A. (Also See Figure 7)

| Name (1) | Grade (2) | $z$ (Quasar) (3) | $z_{\rm phot}$ (Lens) (4) | $g$ (mag) (5) | $r$ (mag) (6) | $z$ (mag) (7) | PXSIG (8) | PMSIG (9) | Avg. Image Sep. (10) |
|---|---|---|---|---|---|---|---|---|---|
| DESI-005.0970-41.7778 | A | 1.58* | | 17.70 | 17.51 | 17.37 | 2.29 | 1.36 | 1.86 |
| DESI-005.0970-41.7778 | | | | 20.98 | 20.49 | 20.72 | 0.19 | 0.54 | |
| DESI-009.3161 + 20.9738 | A | | | 20.59 | 20.18 | 19.62 | | | 1.09 |
| DESI-009.3161 + 20.9738 | | 2.0486 | | 19.45 | 19.25 | 18.99 | 0.01 | 2.29 | |
| DESI-009.4216-07.9179 | A | | | 20.80 | 20.29 | 19.78 | 1.37 | 2.47 | 1.17 |
| DESI-009.4216-07.9179 | | | | 20.86 | 20.38 | 19.82 | 0.38 | 0.45 | |
| DESI-010.3643-02.0590 | A | | | 20.40 | 20.37 | 20.23 | | | 0.57 |
| DESI-010.3643-02.0590 | | | | 19.18 | 19.18 | 19.01 | 0.48 | 4.91 | |
| DESI-014.5397 + 25.4100 | A | 2.60* | | 19.04 | 18.94 | 19.02 | 0.65 | 1.35 | 0.83 |
| DESI-014.5397 + 25.4100 | | | | 20.35 | 19.71 | 19.57 | 0.79 | 1.13 | |
| DESI-018.3481-02.4432 | A | 1.0595 | | 20.69 | 20.54 | 20.56 | | | 0.55 |
| DESI-018.3481-02.4432 | | 1.0595 | | 19.82 | 19.52 | 19.54 | 0.61 | 0.59 | |
| DESI-020.6570-13.7396 | A | | | 20.96 | 20.23 | 19.88 | | | 1.01 |
| DESI-020.6570-13.7396 | | | | 20.27 | 19.86 | 19.33 | 0.91 | 1.70 | |
| DESI-021.3173-10.2081 | A | | 0.420 ± 0.523 | 20.59 | 20.29 | 20.34 | | | 1.13 |
| DESI-021.3173-10.2081 | | | | 19.22 | 19.05 | 19.25 | 0.47 | 1.65 | |
| DESI-023.2696 + 08.2789 | A | | | 20.82 | 20.39 | 20.46 | | | 1.56 |
| DESI-023.2696 + 08.2789 | | | | 20.76 | 20.29 | 20.35 | | | |
| DESI-028.0797-24.8105 | A | 2.42* | 0.286 ± 0.11 | 20.17 | 19.51 | 18.93 | 0.69 | 2.57 | 1.93 |
| DESI-028.0797-24.8105 | | | | 18.73 | 18.65 | 18.49 | 1.03 | 1.51 | |
| DESI-029.1039-27.8562 | A | | | 20.54 | 20.21 | 19.93 | 1.03 | 1.25 | 1.69 |
| DESI-029.1039-27.8562 | | | | 20.48 | 20.15 | 19.87 | 0.45 | 0.50 | |
| DESI-030.0872-15.1609 | A | | | 19.81 | 19.61 | 19.73 | | | 0.80 |
| DESI-030.0872-15.1609 | | | | 20.59 | 20.18 | 19.90 | | | |
| DESI-032.3047-38.6961 | A | | | 20.38 | 20.16 | 20.01 | 1.54 | 1.04 | 1.25 |
| DESI-032.3047-38.6961 | | | | 20.71 | 20.41 | 20.02 | 1.20 | 0.85 | |
| DESI-033.8508-47.4791 | A | | | 20.15 | 19.74 | 19.30 | | | 1.38 |
| DESI-033.8508-47.4791 | | | | 18.99 | 18.72 | 18.50 | 1.99 | 2.02 | |
| DESI-034.1275-55.9702 | A | | | 21.17 | 20.46 | 20.18 | | | 0.81 |
| DESI-034.1275-55.9702 | | | | 19.89 | 19.29 | 19.12 | 2.31 | 2.15 | |
| DESI-035.4474 + 05.9192 | A | 1.52* | | 19.40 | 19.10 | 19.27 | 0.02 | 0.85 | 0.87 |
| DESI-035.4474 + 05.9192 | | | | 20.00 | 19.68 | 19.74 | 0.13 | 0.69 | |
| DESI-037.4925 + 03.3420 | A | | | 18.44 | 18.13 | 18.18 | 0.39 | 0.80 | 2.08 |
| DESI-037.4925 + 03.3420 | | | | 19.05 | 18.66 | 18.61 | 0.42 | 0.58 | |
| DESI-037.5191-07.0795 | A | 2.01* | | 18.03 | 17.68 | 17.41 | | | 0.57 |
| DESI-037.5191-07.0795 | | | | 19.01 | 18.78 | 18.53 | | | |
| DESI-038.0655-24.4942 | A | | | 18.11 | 18.08 | 18.13 | 1.48 | 1.38 | 1.54 |
| DESI-038.0655-24.4942 | | | | 19.62 | 19.38 | 19.13 | 0.92 | 1.35 | |
| DESI-038.1578-58.3572 | A | | | 21.28 | 20.47 | 20.50 | | | 0.73 |
| DESI-038.1578-58.3572 | | | | 18.59 | 18.21 | 17.95 | 0.46 | 2.22 | |
| DESI-040.0767-02.1474 | A | 2.42* | | 19.49 | 19.49 | 19.48 | 0.71 | 0.94 | 0.94 |
| DESI-040.0767-02.1474 | | | | 18.82 | 18.74 | 18.43 | 0.15 | 0.45 | |
| DESI-002.2331-62.9628 | B | | 1.002 ± 0.121 | 21.80 | 21.40 | 21.28 | | | 4.31 |
| DESI-002.2331-62.9628 | | | | 20.93 | 20.63 | 20.55 | 0.16 | 0.34 | |
| DESI-005.4938 + 19.4646 | B | 1.0457 | | 19.88 | 19.79 | 19.63 | 1.24 | 1.99 | 3.06 |
| DESI-005.4938 + 19.4646 | | | | 21.00 | 20.60 | 20.49 | 0.31 | 0.80 | |
| DESI-006.7616 + 02.5372 | B | | | 20.77 | 20.68 | 20.50 | 0.06 | 1.28 | 2.73 |
| DESI-006.7616 + 02.5372 | | | | 20.28 | 20.23 | 20.09 | 0.59 | 0.97 | |
| DESI-007.3861-04.2470 | B | | | 20.73 | 20.22 | 20.01 | | | 1.11 |
| DESI-007.3861-04.2470 | | | | 19.30 | 19.23 | 18.87 | 0.42 | 0.95 | |
| DESI-010.4495-53.8461 | B | | 0.566 ± 0.212 | 21.06 | 20.56 | 20.04 | | | 1.70 |
| DESI-010.4495-53.8461 | | | | 20.97 | 20.54 | 19.82 | 0.67 | 0.92 | |
| DESI-011.3665-39.6262 | B | 2.63* | | 19.63 | 19.69 | 19.32 | 1.31 | 1.14 | 1.33 |
| DESI-011.3665-39.6262 | | | | 18.89 | 19.02 | 18.74 | 0.79 | 0.44 | |
| DESI-011.5839-26.1241 | B | | | 21.47 | 21.45 | 21.05 | | | 2.10 |
| DESI-011.5839-26.1241 | | | | 20.46 | 20.49 | 20.03 | 1.24 | 0.69 | |
| DESI-013.1302-18.9922 | B | | | 21.58 | 21.41 | 21.31 | | | 2.04 |
| DESI-013.1302-18.9922 | | | | 20.82 | 20.60 | 20.58 | | | |
| DESI-014.3211-15.8653 | B | | | 19.11 | 18.98 | 18.78 | 0.25 | 3.58 | 0.82 |
| DESI-014.3211-15.8653 | | | | 20.52 | 20.38 | 20.02 | 0.18 | 2.01 | |
| DESI-015.3368-49.7229 | B | | 0.799 ± 0.157 | 19.68 | 19.52 | 19.29 | 0.56 | 0.80 | 4.05 |
| DESI-015.3368-49.7229 | | | | 20.21 | 19.92 | 19.45 | 0.73 | 0.55 | |
| DESI-016.9070-42.4054 | B | | | 21.29 | 20.87 | 20.19 | 0.14 | 1.96 | 2.63 |

**Table 2**
(Continued)

| Name (1) | Grade (2) | $z$ (Quasar) (3) | $z_{\text{phot}}$ (Lens) (4) | $g$ (mag) (5) | $r$ (mag) (6) | $z$ (mag) (7) | PXSIG (8) | PMSIG (9) | Avg. Image Sep. (10) |
|---|---|---|---|---|---|---|---|---|---|
| DESI-016.9070-42.4054 | | | | 20.20 | 19.93 | 19.61 | 0.39 | 0.95 | |
| DESI-019.6681-31.2618 | B | | | 20.24 | 20.05 | 19.65 | 0.19 | 1.27 | 1.17 |
| DESI-019.6681-31.2618 | | | | 20.09 | 19.46 | 18.83 | 0.44 | 0.38 | |
| DESI-020.5992 + 03.9772 | B | | | 21.04 | 21.01 | 20.85 | | | 1.71 |
| DESI-020.5992 + 03.9772 | | | | 20.60 | 20.51 | 20.27 | 0.32 | 1.00 | |
| DESI-021.0090-63.5788 | B | | | 19.90 | 19.51 | 19.47 | 3.00 | 3.19 | 1.91 |
| DESI-021.0090-63.5788 | | | | 20.47 | 20.28 | 19.96 | 0.98 | 1.92 | |
| DESI-021.1809-35.6025 | B | | | 21.34 | 20.53 | 19.77 | | | 0.59 |
| DESI-021.1809-35.6025 | | | | 21.81 | 20.01 | 19.12 | 1.47 | 2.55 | |
| DESI-022.5859 + 07.4212 | B | 1.5503 | | 19.81 | 19.45 | 19.38 | | | 2.06 |
| DESI-022.5859 + 07.4212 | | | | 19.11 | 18.86 | 18.91 | 0.33 | 0.89 | |
| DESI-025.4316-06.1400 | B | | | 20.41 | 20.28 | 20.12 | | | 1.05 |
| DESI-025.4316-06.1400 | | | | 20.05 | 19.68 | 19.59 | 0.13 | 4.72 | |
| DESI-028.6087-44.0373 | B | | | 20.13 | 20.04 | 19.60 | 1.51 | 1.49 | 4.32 |
| DESI-028.6087-44.0373 | | | | 19.58 | 19.54 | 19.09 | 0.76 | 0.57 | |
| DESI-031.3408 + 10.7791 | B | | | 21.05 | 20.62 | 20.48 | | | 1.36 |
| DESI-031.3408 + 10.7791 | | | | 20.34 | 20.18 | 19.90 | 0.98 | 1.41 | |
| DESI-036.0674-07.6754 | B | | | 21.29 | 21.15 | 20.90 | | | 3.03 |
| DESI-036.0674-07.6754 | | | | 20.70 | 20.56 | 20.28 | | | |
| DESI-038.1065-17.6331 | B | | | 21.14 | 20.04 | 19.52 | | | 0.76 |
| DESI-038.1065-17.6331 | | | | 20.65 | 19.58 | 19.12 | 0.64 | 1.45 | |
| DESI-041.6794-01.5304 | B | | | 20.51 | 19.94 | 19.67 | | | 1.09 |
| DESI-041.6794-01.5304 | | | | 19.14 | 19.03 | 18.65 | 0.26 | 2.55 | |
| DESI-041.8764-26.7729 | B | | | 21.37 | 20.81 | 20.28 | | | 1.23 |
| DESI-041.8764-26.7729 | | | | 20.92 | 20.55 | 20.38 | | | |
| DESI-006.5305 + 19.5864 | C | | | 23.64 | 22.71 | 20.88 | | | 1.27 |
| DESI-006.5305 + 19.5864 | | | | 21.82 | 21.45 | 21.05 | | | |
| DESI-006.9644-47.8883 | C | | | 18.75 | 18.43 | 18.38 | 3.05 | 3.16 | 0.96 |
| DESI-006.9644-47.8883 | | | | 19.37 | 19.04 | 18.66 | 0.06 | 1.50 | |
| DESI-007.2226 + 31.6230 | C | | | 21.76 | 21.40 | 21.31 | | | 1.34 |
| DESI-007.2226 + 31.6230 | | | | 20.58 | 20.40 | 20.10 | 0.72 | 1.08 | |
| DESI-007.5685 + 23.3537 | C | | | 21.26 | 20.91 | 20.78 | | | 0.88 |
| DESI-007.5685 + 23.3537 | | | | 20.93 | 20.54 | 20.42 | 0.87 | 0.97 | |
| DESI-009.3217 + 16.4638 | C | 2.2852 | | 21.21 | 21.01 | 20.81 | | | 2.16 |
| DESI-009.3217 + 16.4638 | | | | 22.33 | 22.08 | 22.07 | | | |
| DESI-009.5577-19.4680 | C | | | 20.89 | 20.70 | 19.87 | | | 1.61 |
| DESI-009.5577-19.4680 | | | | 18.07 | 17.93 | 17.58 | 0.06 | 0.48 | |
| DESI-010.0846-43.3479 | C | | | 22.52 | 22.71 | 22.32 | | | 3.86 |
| DESI-010.0846-43.3479 | | | | 19.21 | 19.12 | 18.96 | 0.66 | 0.51 | |
| DESI-010.9700 + 04.4075 | C | 2.4453 | | 21.26 | 21.20 | 20.93 | | | 2.74 |
| DESI-010.9700 + 04.4075 | | | | 20.07 | 20.14 | 19.85 | 0.31 | 0.65 | |
| DESI-012.0066 + 22.6041 | C | | | 20.91 | 20.71 | 20.44 | | | 2.29 |
| DESI-012.0066 + 22.6041 | | 0.7635 | | 19.16 | 19.33 | 19.06 | 0.19 | 0.47 | |
| DESI-012.2163-56.5312 | C | | | 21.22 | 20.83 | 20.28 | | | 2.35 |
| DESI-012.2163-56.5312 | | | | 20.40 | 20.26 | 19.92 | 0.66 | 0.79 | |
| DESI-012.2718-23.4095 | C | | | 20.74 | 20.59 | 20.55 | 0.30 | 1.15 | 1.98 |
| DESI-012.2718-23.4095 | | | | 20.62 | 20.35 | 20.10 | 0.41 | 0.67 | |
| DESI-012.3029-56.3601 | C | | 0.380 ± 0.208 | 20.67 | 20.40 | 20.24 | 0.24 | 1.26 | 4.97 |
| DESI-012.3029-56.3601 | | | | 19.96 | 19.84 | 19.63 | 0.11 | 0.53 | |
| DESI-012.7346-22.3463 | C | | | 20.50 | 20.49 | 20.03 | 0.16 | 1.13 | 3.69 |
| DESI-012.7346-22.3463 | | | | 18.68 | 18.54 | 18.23 | 0.39 | 0.89 | |
| DESI-016.0726 + 20.6181 | C | | | 20.88 | 20.79 | 20.20 | | | 1.30 |
| DESI-016.0726 + 20.6181 | | | | 19.33 | 19.08 | 18.90 | 1.59 | 0.95 | |
| DESI-017.4428-35.4989 | C | | | 21.40 | 21.05 | 20.88 | | | 0.78 |
| DESI-017.4428-35.4989 | | | | 21.00 | 20.50 | 19.79 | 0.00 | 0.11 | |
| DESI-018.2693 + 00.5923 | C | 0.9436 | | 21.81 | 21.67 | 21.40 | | | 2.00 |
| DESI-018.2693 + 00.5923 | | | | 22.11 | 21.68 | 21.23 | | | |
| DESI-019.5503-01.0784 | C | 0.7395 | | 20.44 | 20.41 | 20.38 | 0.25 | 0.65 | 1.74 |
| DESI-019.5503-01.0784 | | 0.7399 | | 20.11 | 19.77 | 19.31 | 0.37 | 0.58 | |
| DESI-019.6338-41.8640 | C | | | 21.84 | 21.54 | 21.28 | | | 1.90 |
| DESI-019.6338-41.8640 | | | | 21.10 | 20.98 | 20.66 | 0.04 | 0.90 | |
| DESI-022.8138 + 25.1844 | C | | | 20.31 | 20.11 | 19.62 | | | 2.85 |
| DESI-022.8138 + 25.1844 | | | | 20.47 | 20.15 | 19.91 | | | |

**Table 2**
(Continued)

| Name (1) | Grade (2) | $z$ (Quasar) (3) | $z_{\rm phot}$ (Lens) (4) | $g$ (mag) (5) | $r$ (mag) (6) | $z$ (mag) (7) | PXSIG (8) | PMSIG (9) | Avg. Image Sep. (10) |
|---|---|---|---|---|---|---|---|---|---|
| DESI-023.9872-38.5625 | C | | | 22.09 | 21.44 | 20.76 | | | 1.92 |
| DESI-023.9872-38.5625 | | | | 20.64 | 20.33 | 20.40 | 0.53 | 1.38 | |
| DESI-024.0011-06.6026 | C | | 1.247 ± 0.263 | 21.43 | 21.07 | 20.99 | | | 4.54 |
| DESI-024.0011-06.6026 | | | | 21.74 | 21.17 | 21.01 | | | |
| DESI-025.1489-02.8346 | C | | | 22.04 | 21.86 | 21.63 | | | 2.60 |
| DESI-025.1489-02.8346 | | 2.8387 | | 20.49 | 20.27 | 20.07 | 0.05 | 0.71 | |
| DESI-025.4053-22.0813 | C | | | 20.67 | 20.61 | 20.36 | 0.98 | 0.93 | 3.65 |
| DESI-025.4053-22.0813 | | | | 18.72 | 18.60 | 18.44 | 0.42 | 0.50 | |
| DESI-027.1003 + 02.0790 | C | | 0.269 ± 0.094 | 21.30 | 20.57 | 20.45 | | | 4.79 |
| DESI-027.1003 + 02.0790 | | | | 22.62 | 22.36 | 22.40 | | | |
| DESI-033.3456-04.3595 | C | 3.5458 | | 20.98 | 20.79 | 20.66 | | | 1.91 |
| DESI-033.3456-04.3595 | | 1.9052 | | 19.38 | 19.26 | 19.02 | 0.09 | 1.19 | |
| DESI-033.5044-21.7731 | C | | | 20.59 | 20.52 | 20.24 | | | 1.35 |
| DESI-033.5044-21.7731 | | | | 20.05 | 19.84 | 19.72 | 0.42 | 1.21 | |
| DESI-034.1526 + 14.5129 | C | | | 20.27 | 19.77 | 19.21 | | | 0.56 |
| DESI-034.1526 + 14.5129 | | | | 20.10 | 19.77 | 19.73 | 0.38 | 1.36 | |
| DESI-034.2775 + 15.3661 | C | | | 23.04 | 21.77 | 20.77 | | | 1.37 |
| DESI-034.2775 + 15.3661 | | | | 22.29 | 21.68 | 21.46 | | | |
| DESI-035.3347-29.5103 | C | | | 21.96 | 20.29 | 19.45 | 0.33 | 0.46 | 1.18 |
| DESI-035.3347-29.5103 | | | | 19.28 | 18.59 | 18.40 | 1.41 | 0.18 | |
| DESI-035.5218-23.6958 | C | | | 20.61 | 20.46 | 20.32 | 0.70 | 0.77 | 2.11 |
| DESI-035.5218-23.6958 | | | | 18.94 | 18.98 | 19.07 | 0.34 | 0.52 | |
| DESI-035.9851-42.8070 | C | | | 19.50 | 18.93 | 18.64 | 0.00 | 1.40 | 1.81 |
| DESI-035.9851-42.8070 | | | | 18.71 | 18.64 | 18.68 | 0.53 | 0.84 | |
| DESI-037.7436-02.1156 | C | 1.7392 | | 21.29 | 21.15 | 20.95 | | | 3.25 |
| DESI-037.7436-02.1156 | | | | 19.67 | 19.48 | 19.23 | 0.56 | 1.11 | |
| DESI-038.1300 + 17.9931 | C | | | 18.84 | 18.27 | 17.87 | 0.47 | 2.78 | 1.56 |
| DESI-038.1300 + 17.9931 | | | | 20.81 | 19.58 | 19.12 | 0.27 | 1.62 | |
| DESI-038.6539-50.1670 | C | | | 21.00 | 19.77 | 19.26 | 1.20 | 1.45 | 2.34 |
| DESI-038.6539-50.1670 | | | | 19.86 | 19.31 | 18.74 | 0.51 | 0.20 | |
| DESI-040.6886-10.0492 | C | | | 19.63 | 19.49 | 19.27 | 0.11 | 1.61 | 2.39 |
| DESI-040.6886-10.0492 | | | | 19.19 | 18.69 | 18.35 | 0.54 | 0.22 | |
| DESI-041.7404-34.6027 | C | | | 20.92 | 20.66 | 20.60 | 0.21 | 1.03 | 4.21 |
| DESI-041.7404-34.6027 | | | | 19.33 | 19.17 | 19.00 | 0.40 | 0.82 | |

**Notes.** The first 80 new candidate systems arranged in ascending R.A., all of which are doubles. Thus, each system has two rows, corresponding to the two quasar images. To avoid confusion, the two rows for each system have the same system name. In the table on our project website, we provide a cutout centered on the individual quasar image in each system. Therefore, there will be no ambiguity. The columns are arranged as follows. Column (1): name of the system (R.A. and decl. in decimal format). Column (2): human inspection grade. Column (3): spectroscopic redshift of quasars from SDSS DR16 or SNIFS (with an asterisk). Column (4): photometric redshift of the putative lenses from Zhou et al. (2020). Columns (5), (6), and (7): the $g$-, $r$-, and $z$-band magnitudes, respectively. Column (8): parallax significance. Column (9): PM significance. Column (10) average separation of images in arcseconds. This portion of the table only shows the first 80 new candidates, while the full online version has the complete set of 436 candidates.

(This table is available in its entirety in machine-readable form.)

image separation 3.″69.[17] Last, we have found a potential double-source lensing system, DESI-181.2111 + 44.4764. The two quasar images have SDSS redshifts of 1.1438 and 1.8095, respectively, with an image separation of 3.″99. In between these two images, there is clearly a galaxy foreground to both quasar images (based on color), which we consider to be the putative lens.

For 11 of the 18 targets shown in Figure 9, the S/N is sufficiently high to see plausible evidence of multiple sources in wavelength slices from SNIFS data cubes that correspond to emission features. In Figure 10, we show wavelength slices for these emission features: Ly$\alpha$ (1215 Å), Si IV+O IV (1400 Å), C IV (1549 Å), [C III] (1909 Å), Mg II (2799 Å), H$\beta$ (4863 Å), and the [O III] doublet (4960 Å, 5008 Å). While the last two features (H$\beta$ and the [O III] doublet) can be observed for both quasars and galaxies, the first five are all features much more likely to be present for a quasar than a galaxy.[18] Ten of the 11 systems have three to five of the five features predominantly seen in quasars, as mentioned above. For the last system, it is true that the emission features of H$\beta$ and the [O III] doublet could be due to the quasar and its host galaxy, but it is still more plausible that the features are from multiple quasar images, because the contours are drawn around objects from a quasar catalog. Each slice is 6.″4 on the side, corresponding to the SNIFS field of view. The shifting of the emission features from the longest to the shortest wavelength is due to atmospheric differential refraction (ADR). The ADR effect is greater for images further south, as expected, given the higher airmass from Maunakea. These wavelength slices show that in these 11 systems, multiple quasar images are at the same redshift, suggesting that they are either lensing systems or binary quasars.

[17] It is also possible that these candidates are merely projected close pairs, but the background quasar is not strongly lensed, in which case they are still useful as "quasars probing quasars" systems (e.g., Findlay et al. 2018).

[18] E.g., https://classic.sdss.org/dr6/algorithms/linestable.php.

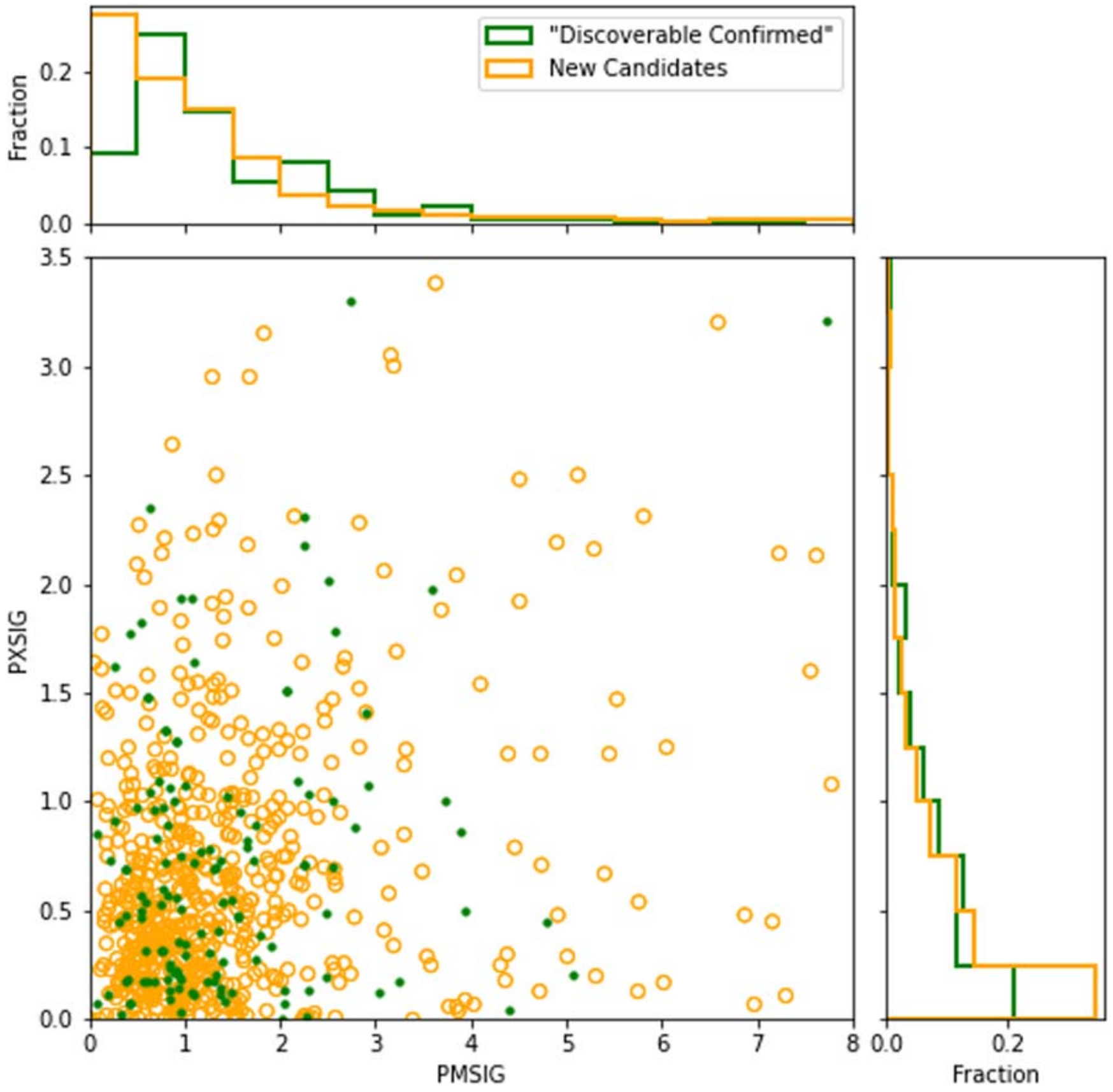

**Figure 8.** The distributions of significance of both PM and parallax in the images of our newly discovered candidates and of "discoverable confirmed" systems (see Table 1).

The SNIFS pilot study shows that to *fully* confirm a high fraction of these candidate systems, deeper observations with higher resolution are needed. Adaptive-optics-assisted integral field spectrographs (e.g., Very Large Telescope MUSE and Keck OSIRIS) would be particularly useful to obtain redshifts of the lens and multiple lensed images. Such observations would be also needed for lens modeling and constraining $H_0$.

#### *4.2.3. Candidates with Both Foreground and Background Redshifts*

The following six candidate systems have photometric redshifts for the foreground galaxy and spectroscopic redshifts for the background quasar, with three Grade As, two Bs, and one C. We note that the lensing galaxies are ellipticals, which typically have reliable photometric redshifts (Zhou et al. 2020).[19] The redshifts and image separations for these systems are consistent with lensing. However, for full confirmation, spectroscopic and high-resolution imaging and/or modeling are needed. We provide a brief description for each of them below.

DESI-028.0797-24.8105: this Grade A candidate is one of the visually impressive systems shown in Figure 6. It is also one of the first 80 systems arranged by R.A. (Table 2). It was observed by SNIFS (Figure 9) and we determined the quasar redshift to be 2.42. The photometric redshift for the foreground galaxy, the putative lens, is $0.286 \pm 0.110$. One of the two quasar images has the second highest PMSIG among the quasar images for these six systems, at 2.57, but it has a low PXSIG of 0.69. The quasar image separation is $1\farcs93$.

DESI-060.4504-25.2439: this Grade A system is also one of the visually impressive systems shown in Figure 6. Its redshift was determined to be 1.33 from SNIFS observation (Figure 9). The putative lensing galaxy has a photometric redshift of $0.465 \pm 0.159$. All images have relatively low PXSIG and PMSIG, consistent with zero. The quasar image separation is $1\farcs45$.

DESI-251.5695 + 44.0197: this is the third of the three Grade A systems. SDSS provides a quasar redshift of 2.0757, and the photometric redshift for the foreground galaxy is $0.620 \pm 0.048$. We did not find Gaia parallax or PM information for the images in this system. The image separation is $3\farcs43$.

DESI-141.7488 + 06.3905: this is a Grade B system. The SDSS redshift for one quasar image is 0.6671 and the foreground galaxy has $z_{\rm phot} = 0.620 \pm 0.048$. One of the images has the highest PMSIG among the quasar images for these six systems, at 3.87, but the PXSIG is only 0.04. This candidate has the largest image separation among the six systems, at $4\farcs96$.

DESI-183.7095 + 07.8064: this is also a Grade B system. One of the quasar images and the foreground galaxy have a redshift of 1.6920 from SDSS and a photometric redshift of $0.622 \pm 0.060$, respectively. No Gaia data are available. The image separation is 4.″78.

[19] Note that as *The Tractor* is a forward-modeling source extraction algorithm, it provides a model for both the quasar images and the putative lens, so the photometry for successfully extracted objects (such as the lensing galaxy) should still be fairly reliable.

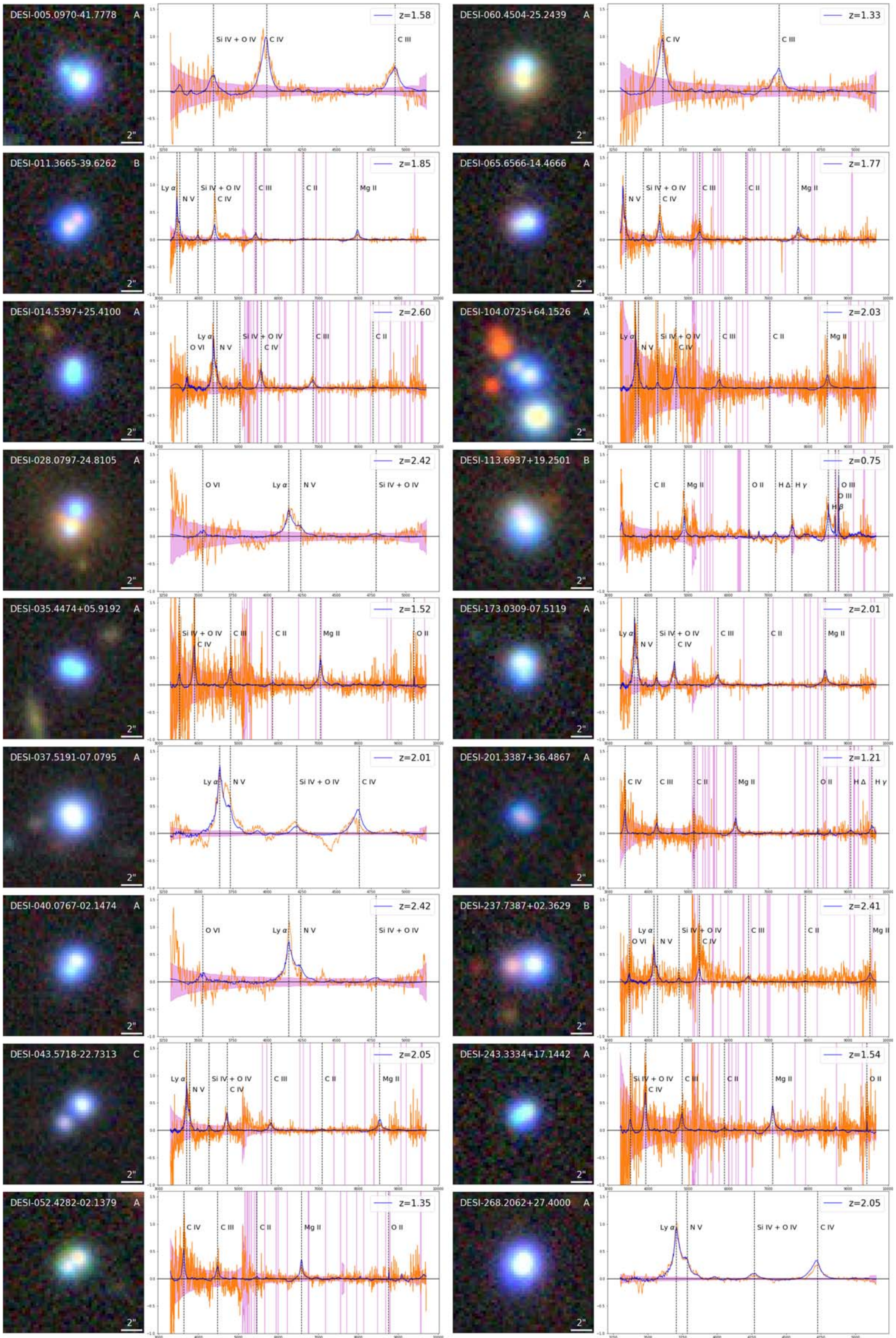

**Figure 9.** Legacy Surveys image cutouts alongside spectra from SNIFS on UH88 for 18 observed candidates with sufficient S/N ratio. We perform a correlation between the spectra and the SDSS quasar spectral template. Here we report the best-matched redshift. For each system, the spectrum is shown in orange, with the error spectrum in magenta shading. Redshifted templates are shown in blue, with the black lines representing emission features. DESI-237.7387 + 02.3629 was previously observed by SDSS with consistent redshift. Note that, for some systems, only the blue channel of SNIFS was available at the time of observation.

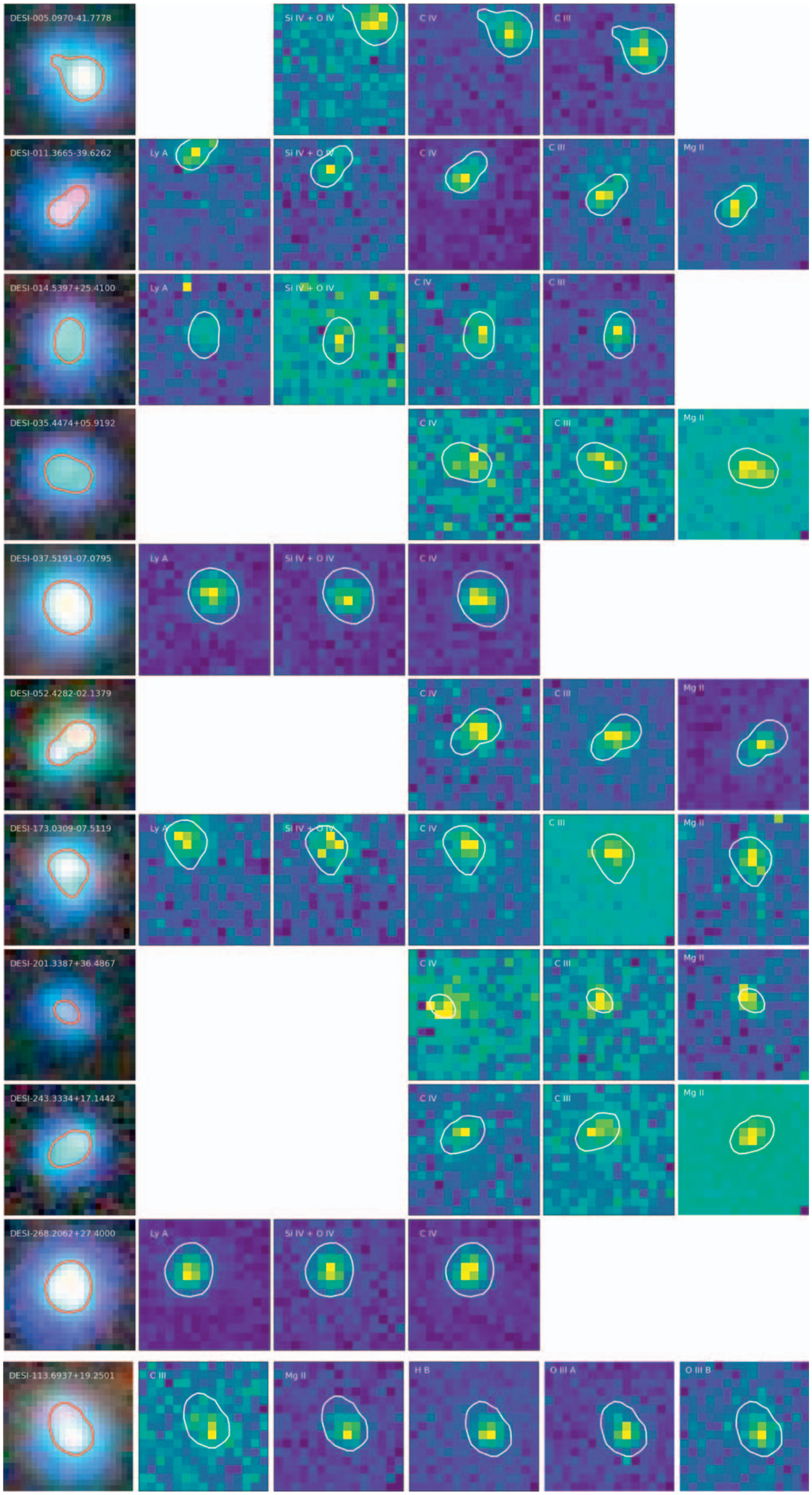

**Figure 10.** Each column shows an observed wavelength slice from SNIFS, corresponding to emission features in ascending wavelength, from Ly$\alpha$ to Mg II (see the text). We impose the contour from the Legacy Survey images (set to 80% of the brightest pixel for each image) on each of the slices. For each system, the extent, shape, and orientation of the contour matches the brightest pixels in the SNIFS slices, suggesting the presence of two or more source images. For each of these 11 systems (one per row), these wavelength slices show that there are multiple quasar images at the same redshift. The emission features of the last row do not align with the rest, given its lower redshift. Note the effects of the ADR (see the text).

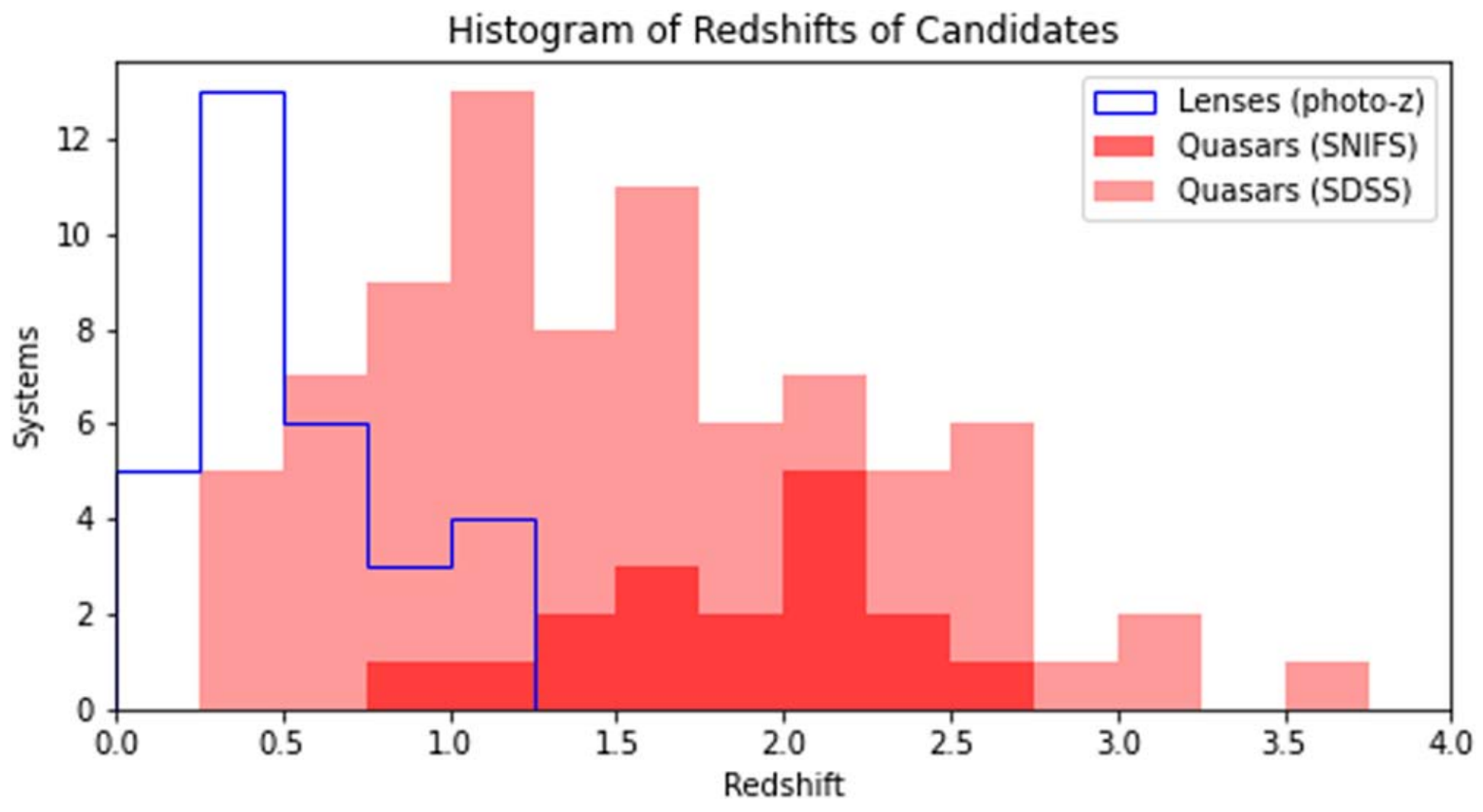

**Figure 11.** The redshift distributions for the lenses and quasars of our candidate systems. The redshifts of our quasars are from SDSS DR16 and, for 17 additional systems, from the SNIFS instrument on the University of Hawaii 88 inch (2.2 m) telescope. Photometric redshifts for the lenses are from Zhou et al. (2020).

DESI-129.2063 + 48.6978: this is the only Grade C system. A quasar image and the foreground galaxy have a redshift of 1.7107 from SDSS and a photometric redshift of $0.776 \pm 0.245$, respectively. The Gaia PM and parallax measurements are consistent with zero. The two quasar images are separated by 4.″22.

## 5. Discussion

In this section, we assess the fraction of our Grade A candidates that are confirmed as quasars (Section 5.1) and compare the double-to-quad ratio for our candidates with forecast results (Section 5.2).

### 5.1. Contamination Estimation

For SNIFS targets, we selected visually convincing systems with $r$-band mag $\lesssim$20.0, which approximately corresponds to the brighter half of the distribution (Figure 4). Not surprisingly, a strong majority of these targets are Grade A systems: 15 of the 19 targets with sufficient S/N. One of them turned out to be a star (as mentioned in Section 4.2.2, this was subsequently removed from our candidate list), with the rest confirmed to be quasars. Among the 102 Grade A candidates, as with the rest of our candidates, approximately half of them (55) have at least one image with $r$-band mag $<$20.0. Based on the SNIFS targets, we expect 93% of them, or 51, to be actual quasars.

### 5.2. Double-to-quad Ratio

For the 436 new lensed quasar candidates discovered in the Legacy Surveys using the DESI Quasar Sample plus the 94 "discoverable known" systems from the same sample, 506 are doubles and 24 are quads. This amounts to a double/quad ratio of ∼21 to 1, much higher than expected. Oguri & Marshall (2010; hereafter, OM10), while estimating the double/quad ratio of detectable lensed quasars (those with a minimum image separation of $(2/3)^{*}\theta_{\rm PSF}$) in various surveys according to their depth, predict that the DES will contain a double/quad ratio of ∼6 to 1 (they specifically cite the percentage of quads). DES is part of the DECaLS subregion of the Legacy Surveys. By filtering our candidates appropriately, we may more fairly compare our double/quad ratio to that estimated in DES by OM10.

In order to compare with the OM10 prediction, we restrict our sample as follows:

1. The MzLS/BASS subregion has worse seeing in the $gr$ bands than DECaLS. Given our search algorithm, this will have a significant effect on the double/quad ratio in MzLS/BASS compared to DECaLS. Because most quads are not perfect Einstein crosses and are instead asymmetrical, often with three close-by (even blended) images and one fainter counterimage, this would imply only two objects would be identified by the DESI Quasar Sample. Our algorithm would then identify them as a double, resulting in the loss of a disproportionate number of quads. Thus, for comparison with the OM10 ratio, we consider the double/quad ratio for our candidates in DECaLS only.
2. OM10 only considered systems with image separation $>(2/3)^{*}\theta_{\rm PSF}$, so using $\theta_{\rm PSF} = 1.''18$ in the $r$ band in DECaLS, we further restrict our candidates in DECaLS to those with image separation >0.″79.
3. We also restrict our candidates to those graded as As. This decision is motivated primarily by our higher confidence in As compared to Bs and Cs. While we are confident in the quality of our B- and C-grade candidates, we are less confident in their typing as doubles or quads.

With the above restrictions—that is, A-grade candidates with image separation >0.″79 within the DECaLS footprint—our double/quad ratio is ∼11.4 to 1 (80 to 7), which still differs from that predicted for DES (∼6 to 1). Poisson noise alone does not seem to fully account for the discrepancy between our ratio and OM10's. One possible explanation for the discrepancy is that some of the doubles are physical quasar–quasar binaries, rather than doubly lensed quasars (e.g., Lemon et al. 2020). We note in Section 1 that physical binaries are also important for quasar physics. Follow-up spectroscopic and high-resolution observations are needed to determine the fraction of our candidates that are physical binaries.

Traditionally, quads have been more highly valued because they more tightly constrain lens modeling (Suyu et al. 2017). Doubles, though, generally have much more precise time-delay measurements, because their time delays tend to be longer than those of quads (OM10). Current measurements of $H_0$ using doubles are becoming more competitive. For example, using 27 doubly lensed quasars, taking into account various systematic effects, Harvey (2020) reported a 4% measurement of $H_0$. This is important, as doubles will dominate over quads among the lensed quasars yet to be discovered. Magnification bias strongly favors quads (Wallington & Narayan 1993), so deeper searches for lensed quasars are expected to yield more doubles than quads (e.g., Agnello et al. 2018; Lemon et al. 2019, 2020; Chan et al. 2020). With future surveys slated to produce ever higher double/quad ratios—the Vera C. Rubin Legacy Survey of Space and Time is expected to deliver a double/quad ratio of ∼7 to 1 (OM10)—doubly lensed quasars can play a significant role toward 1% precision $H_0$ measurements.

## 6. Conclusions

We have carried out a search for multiply lensed quasar systems in the DESI Legacy Imaging Surveys. We first apply an autocorrelation algorithm to the 5 million targets in the DESI Quasar Sample (Yèche et al. 2020). Then, guided by the visual appearance of the confirmed systems in the Legacy Surveys footprint, we inspect ∼6000 systems with image separations $<5''$ found by our algorithm. Among these, we discover and report 436 new lensed quasar candidates, with 102 Grade As, 118 Bs, and 216 Cs. We have found quasar redshifts for 65 systems from SDSS DR16, which range from 0.36 to 3.55. Of these, 61 have $z > 0.5$. We have obtained spectra using SNIFS on the University of Hawaii 2.2 m telescope for 18 additional quasars (one of which was previously observed by SDSS with consistent redshift). The best-fit redshifts for these 18 range from $z = 0.75$ to 2.60. The high-redshift distribution of the quasars is consistent with them being lensed. Based on the SNIFS observations, we estimate ∼93% of the brighter half of our Grade A candidates are actual quasars. Since our candidates are discovered from the DESI Quasar Sample, all of them will be spectroscopically observed by DESI. Among our candidates, we have identified the following: one system with BAL activity; three potential rare quasar–quasar lensing systems; a possible double-source lensing system; and six candidates with both foreground and background redshifts that are consistent with lensing.

These are very promising findings and represent a significant addition to the existing sample of lensed quasars. Oguri & Marshall (2010) predicted a double/quad ratio in DES of 6 to 1. After appropriate restrictions are applied, the corresponding ratio for our search results is ∼11.4 to 1. Even after accounting for Poisson noise, it is possible that some of the doubles among our candidates are physical binaries, which are important for gaining a deeper understanding of quasar physics. These discoveries will contribute a large number of systems for the time-delay measurement of $H_0$ with high accuracy and precision.

## Acknowledgments

This work was supported in part by the Director, Office of Science, Office of High Energy Physics of the US Department of Energy, under contract No. DE-AC025CH11231. This research used resources of the National Energy Research Scientific Computing Center (NERSC), a U.S. Department of Energy Office of Science User Facility operated under the same contract as above, and the Computational HEP program in The Department of Energy's Science Office of High Energy Physics provided resources through the Cosmology Data Repository project (grant No. KA2401022). This work was also supported in part by the U.S. Department of Energy, Office of Science, Office of Workforce Development for Teachers and Scientists (WDTS) under the Science Undergraduate Laboratory Internship (SULI) program. C.D. is thankful for the support from the Office of Undergraduate Research Student Initiated Internship Program (OURSIP) at Princeton University. X.H. acknowledges the University of San Francisco Faculty Development Fund. We thank the technical staff of the University of Hawaii 2.2 m telescope. We recognize the significant cultural role of Maunakea within the indigenous Hawaiian community, and we appreciate the opportunity to conduct observations from this revered site.

The Legacy Surveys consist of three individual and complementary projects: the Dark Energy Camera Legacy Survey (DECaLS; Proposal ID #2014B-0404; PIs: David Schlegel and Arjun Dey), the Beijing–Arizona Sky Survey (BASS; NOAO Prop. ID #2015A-0801; PIs: Zhou Xu and Xiaohui Fan), and the Mayall $z$-band Legacy Survey (MzLS; Prop. ID #2016A-0453; PI: Arjun Dey). DECaLS, BASS, and MzLS together include data obtained, respectively, at the Blanco telescope, Cerro Tololo Inter-American Observatory, NSF's NOIRLab; the Bok telescope, Steward Observatory, University of Arizona; and the Mayall telescope, Kitt Peak National Observatory, NOIRLab. Pipeline processing and analyses of the data were supported by NOIRLab and the Lawrence Berkeley National Laboratory (LBNL). The Legacy Surveys project is honored to be permitted to conduct astronomical research on Iolkam Du'ag (Kitt Peak), a mountain with particular significance to the Tohono O'odham Nation. NOIRLab is operated by the Association of Universities for Research in Astronomy (AURA) under a cooperative agreement with the National Science Foundation. LBNL is managed by the Regents of the University of California under contract to the U.S. Department of Energy. This project used data obtained with the Dark Energy Camera (DECam), which was constructed by the Dark Energy Survey (DES) collaboration. Funding for the DES Projects has been provided by the U.S. Department of Energy, the U.S. National Science Foundation, the Ministry of Science and Education of Spain, the Science and Technology Facilities Council of the United Kingdom, the Higher Education Funding Council for England, the National Center for Supercomputing Applications at the University of Illinois at Urbana-Champaign, the Kavli Institute of Cosmological Physics at the University of Chicago, the Center for Cosmology and Astro-Particle Physics at the Ohio State University, the Mitchell Institute for Fundamental Physics and Astronomy at Texas A&M University, Financiadora de Estudos e Projetos, Fundação Carlos Chagas Filho de Amparo à Pesquisa do Estado do Rio de Janeiro, Conselho Nacional de Desenvolvimento Científico e Tecnológico and the Ministério da Ciência, Tecnologia e Inovacão, the Deutsche Forschungsgemeinschaft, and the Collaborating Institutions in the Dark Energy Survey. The Collaborating Institutions are Argonne National Laboratory, the University of California at Santa Cruz, the University of Cambridge, Centro de Investigaciones Enérgeticas, Medioambientales y Tecnológicas-Madrid, the University of Chicago, University College London, the DES–Brazil Consortium, the University of Edinburgh, the Eidgenössische Technische Hochschule (ETH) Zürich, Fermi National Accelerator Laboratory, the University of Illinois at Urbana-Champaign, the Institut de Ciències de l'Espai (IEEC/CSIC), the Institut de Física d'Altes Energies, Lawrence Berkeley National Laboratory, the Ludwig-Maximilians Universität München and the associated Excellence Cluster Universe, the University of Michigan, NSF's NOIRLab, the University of Nottingham, the Ohio State University, the University of Pennsylvania, the University of Portsmouth, SLAC National Accelerator Laboratory, Stanford University, the University of Sussex, and Texas A&M University. BASS is a key project of the Telescope Access Program (TAP), which has been funded by the National Astronomical Observatories of China, the Chinese Academy of Sciences (the Strategic Priority Research Program The Emergence of Cosmological Structures, grant No. XDB09000000), and the Special Fund for Astronomy from the Ministry of Finance. BASS is also supported by the External Cooperation Program of the Chinese Academy of Sciences (grant No. 114A11KYSB20160057) and the Chinese National Natural Science Foundation (grant Nos. 12120101003 and 11433005). The Legacy Survey team makes use of data products from the Near-Earth Object Wide-field Infrared Survey Explorer (NEOWISE), which is a project of the Jet Propulsion Laboratory/California Institute of Technology. NEOWISE is funded by the National Aeronautics and Space Administration. The Legacy Surveys imaging of the DESI footprint is supported by the Director, Office of Science, Office of High Energy Physics of the U.S. Department of Energy, under Contract No. DE-AC02-05CH1123; by the National Energy Research Scientific Computing Center, a DOE Office of Science User Facility, under the same contract; and by the U.S. National Science Foundation, Division of Astronomical Sciences, under Contract No. AST-0950945 to NOAO.

## ORCID iDs

C. Dawes https://orcid.org/0000-0002-3367-6053
C. Storfer https://orcid.org/0000-0002-0385-0014
X. Huang https://orcid.org/0000-0001-8156-0330
Aleksandar Cikota https://orcid.org/0000-0001-7101-9831
Arjun Dey https://orcid.org/0000-0002-4928-4003
D. J. Schlegel https://orcid.org/0000-0002-5042-5088

## References

Agnello, A., Lin, H., Kuropatkin, N., et al. 2018, MNRAS, 479, 4345
Aldering, G., Adam, G., Antilogus, P., et al. 2002, Proc. SPIE, 4836, 61
Anguita, T., Schechter, P. L., Kuropatkin, N., et al. 2018, MNRAS, 480, 5017
Begelman, M. C., Blandford, R. D., & Rees, M. J. 1980, Natur, 287, 307
Birrer, S., Shajib, A. J., Galan, A., et al. 2020, A&A, 643, A165
Birrer, S., & Treu, T. 2021, A&A, 649, A61
Bogdanović, T., Miller, M. C., & Blecha, L. 2022, LRR, 25, 3
Boroson, T. A., & Lauer, T. R. 2009, Natur, 458, 53
Chan, J. H. H., Suyu, S. H., Sonnenfeld, A., et al. 2020, A&A, 636, A87
Delchambre, L., Krone-Martins, A., Wertz, O., et al. 2019, A&A, 622, A165
Dey, A., Schlegel, D. J., Lang, D., et al. 2019, AJ, 157, 168
Di Matteo, T., Springel, V., & Hernquist, L. 2005, Natur, 433, 604
Ding, X., Treu, T., Birrer, S., et al. 2021, MNRAS, 501, 269
Ding, X., Treu, T., Suyu, S. H., et al. 2017, MNRAS, 472, 90
Djorgovski, S. 1991, in ASP Conf. Ser. 21, The Space Distribution of Quasars, ed. D. Crampton (San Francisco, CA: ASP), 349
Ellison, S. L., Patton, D. R., Mendel, J. T., & Scudder, J. M. 2011, MNRAS, 418, 2043
Ferrarese, L., & Merritt, D. 2000, ApJL, 539, L9
Findlay, J. R., Prochaska, J. X., Hennawi, J. F., et al. 2018, ApJS, 236, 44
Gebhardt, K., Bender, R., Bower, G., et al. 2000, ApJL, 539, L13
Häring, N., & Rix, H.-W. 2004, ApJL, 604, L89
Harvey, D. 2020, MNRAS, 498, 2871
Hopkins, P. F., Hernquist, L., Cox, T. J., & Kereš, D. 2008, ApJS, 175, 356
Huang, X., Storfer, C., Gu, A., et al. 2021, ApJ, 909, 27
Huang, X., Storfer, C., Ravi, V., et al. 2020, ApJ, 894, 78
Inada, N., Becker, R. H., Burles, S., et al. 2003, AJ, 126, 666
Inada, N., Oguri, M., Shin, M.-S., et al. 2012, AJ, 143, 119
Jaelani, A. T., Rusu, C. E., Kayo, I., et al. 2021, MNRAS, 502, 1487
Kochanek, C. S., Falco, E. E., & Muñoz, J. A. 1999, ApJ, 510, 590
Kormendy, J., & Richstone, D. 1995, ARA&A, 33, 581
Lang, D., Hogg, D. W., & Mykytyn, D., 2016 The Tractor: Probabilistic astronomical source detection and measurement, Astrophysics Source Code Library, ascl:1604.008
Lantz, B., Aldering, G., Antilogus, P., et al. 2004, Proc. SPIE, 5249, 146
Lemon, C., Auger, M. W., McMahon, R., et al. 2020, MNRAS, 494, 3491
Lemon, C. A., Auger, M. W., & McMahon, R. G. 2019, MNRAS, 483, 4242
Lemon, C. A., Auger, M. W., McMahon, R. G., & Ostrovski, F. 2018, MNRAS, 479, 5060
Liu, X., Shen, Y., & Strauss, M. A. 2012, ApJ, 745, 94
Magorrian, J., Tremaine, S., Richstone, D., et al. 1998, AJ, 115, 2285
Marconi, A., & Hunt, L. K. 2003, ApJL, 589, L21
More, A., Oguri, M., Kayo, I., et al. 2016, MNRAS, 456, 1595
Mortlock, D. J., Webster, R. L., & Francis, P. J. 1999, MNRAS, 309, 836
Moustakas, L. A., Brownstein, J., Fadely, R., et al. 2012, AAS Meeting, 219, 146.01
Oguri, M., & Marshall, P. J. 2010, MNRAS, 405, 2579
Planck Collaboration, Aghanim, N., Akrami, Y., et al. 2020, A&A, 641, A6
Rathna Kumar, S., Stalin, C. S., & Prabhu, T. P. 2015, A&A, 580, A38
Refsdal, S. 1964, MNRAS, 128, 307
Riess, A. G., Casertano, S., Yuan, W., Macri, L. M., & Scolnic, D. 2019, ApJ, 876, 85
Riess, A. G., Yuan, W., Macri, L. M., et al. 2022, ApJL, 934, L7
Rusu, C. E., Berghea, C. T., Fassnacht, C. D., et al. 2019, MNRAS, 486, 4987
Sergeyev, A. V., Zheleznyak, A. P., Shalyapin, V. N., & Goicoechea, L. J. 2016, MNRAS, 456, 1948
Shajib, A. J., Birrer, S., Treu, T., et al. 2020, MNRAS, 494, 6072
Sonnenfeld, A., Chan, J. H. H., Shu, Y., et al. 2018, PASJ, 70, S29
Sonnenfeld, A., Verma, A., More, A., et al. 2020, A&A, 642, A148
Spiniello, C., Agnello, A., Napolitano, N. R., et al. 2018, MNRAS, 480, 1163
Stein, G., Blaum, J., Harrington, P., Medan, T., & Lukić, Z. 2022, ApJ, 932, 107
Stern, D., Djorgovski, S. G., Krone-Martins, A., et al. 2021, ApJ, 921, 42
Storfer, C., Huang, X., Gu, A., et al. 2022, arXiv:2206.02764
Suyu, S. H., Bonvin, V., Courbin, F., et al. 2017, MNRAS, 468, 2590
Treu, T., & Marshall, P. J. 2016, A&ARv, 24, 11
Wallington, S., & Narayan, R. 1993, ApJ, 403, 517
Weymann, R. J., Chaffee, F. H., & Davis, J. 1979, ApJL, 233, L43
Wong, K. C., Suyu, S. H., Chen, G. C. F., et al. 2020, MNRAS, 498, 1420
Wright, E. L., Eisenhardt, P. R. M., Mainzer, A. K., et al. 2010, AJ, 140, 1868
Yèche, C., Palanque-Delabrouille, N., Claveau, C.-A., et al. 2020, RNAAS, 4, 179
Zhou, R., Newman, J. A., Mao, Y.-Y., et al. 2021, MNRAS, 501, 3309
Zou, H., Zhou, X., Fan, X., et al. 2017, PASP, 129, 064101